\documentclass[aps,prc,showpacs,showkeys,superscriptaddress,nofootinbib,twocolumn,floatfix]{revtex4}

\usepackage{bm}
\usepackage{placeins}
\usepackage{graphicx}
\usepackage{amsfonts}
\usepackage{bbm}
\usepackage{multirow}
\usepackage{graphicx,color,amsmath,amssymb}

\usepackage{hyperref}
\hypersetup{
            colorlinks=true,
            linkcolor=blue,
            anchorcolor=blue,
            citecolor=blue}

\def\eg{{\em e.g.}}
\def\ie{{\em i.e.}}

\def\Meff{M_{\rm eff}}
\def\Tpc{T_{\rm pc}}
\newcommand{\beq}{\begin{equation}}
\newcommand{\eeq}{\end{equation}}
\newcommand{\bea}{\begin{eqnarray}}
\newcommand{\eea}{\end{eqnarray}}

\begin{document}

\title{Bottomonium Properties in QGP from a Lattice-QCD Informed $T$-Matrix Approach}

\author{Zhanduo Tang}
\affiliation{Cyclotron Institute and Department of Physics and Astronomy, Texas A\&M University, College Station, TX 77843-3366, USA}

\author{Swagato Mukherjee}
\author{Peter Petreczky}
\affiliation{Physics Department, Brookhaven National Laboratory, Upton, New York 11973, USA}

\author{Ralf Rapp}
\affiliation{Cyclotron Institute and Department of Physics and Astronomy, Texas A\&M University, College Station, TX 77843-3366, USA}

\begin{abstract}
Recent computations of bottomonium correlation functions with extended sources in lattice-discretized Quantum Chromodynamics (lQCD) provide new insights into heavy-quark dynamics at distance scales which are of the order of the inverse temperature. We analyze these results employing the thermodynamic $T$-matrix framework, in a continued effort to interpret lQCD data for quarkonium correlation functions in a non-perturbative and selfconsistently solved quantum-many body approach to a strongly coupled quark-gluon plasma (QGP). 
Its key inputs are the in-medium driving kernel (potential) of the scattering equation and an interference function  which implements 3-body effects in the quarkonium coupling to the thermal medium.
A simultaneous description of lQCD results for the bottomonium correlators with extended operators and the previously analyzed Wilson line correlators
only requires minor refinements of the potential but calls for stronger
interference effects at larger separation of the bottom quark and antiquark.
We then analyze the poles of the selfconsistent $T$-matrices on the real axis to assess the survival of the various bound states. We estimate the pertinent temperatures where the poles disappear for
the various bottomonium states and discuss the relation to the 
corresponding peaks in the bottomonium spectral functions.
We also recalculate the spatial diffusion coefficient of the QGP and find it to be similar to that in our previous study.
\end{abstract}

\maketitle


\section{Introduction} 
\label{sec_intro} 
The microscopic description of the quark-gluon plasma (QGP) as emerging from the underlying interactions of Quantum Chromodynamics (QCD) is a formidable challenge. 
Heavy-flavor (HF) particles, \ie, charm (c) and bottom (b) quarks, are regarded as valuable probes in this respect as their large masses enable approximations
that render the resulting quantum many-body problem much more tractable. In addition, HF particles offer several attractive features to study the QCD medium in ultra-relativistic heavy-ion collisions (URHICs)~\cite{Prino:2016cni,Dong:2019unq}: heavy quarks are produced in initial hard processes and their numbers are approximately conserved throughout the fireball's evolution; their thermal relaxation time is prolonged compared to that of light partons and similar to, or even larger than, the fireball lifetime, thereby preserving a memory of their interaction history. 
The Brownian motion of low-momentum heavy quarks has therefore developed into a prime mean to extract their spatial diffusion coefficient which is a fundamental transport coefficient of the QGP.

The study of heavy quarkonia, \ie, bound states of a heavy quark ($Q$) and antiquark ($\bar{Q}$), is, in principle, more directly related to the in-medium properties of the QCD force. However, while the pertinent observables in URHICs, \eg,  quarkonia abundances and transverse-momentum spectra, largely depend on their in-medium properties (in particular in-medium widths and binding energies), the interpretation of the data is usually more involved~\cite{Rapp:2008tf,Kluberg:2009wc,Braun-Munzinger:2009dzl,Andronic:2024oxz} than in the open HF sector.
Yet, it is important to note that the microscopic processes underlying the medium effects in the open and hidden HF sectors are closely related (\eg, the heavy-light interactions that drive heavy-quark (HQ) diffusion in the QGP are a key ingredient to compute the quarkonium dissociation widths). 

In the present study we focus on bottomonia whose in-medium properties have been extensively explored in both lattice QCD~\cite{Aarts:2013kaa,Aarts:2014cda,Kim:2014iga,Larsen:2019bwy,Larsen:2019zqv,Larsen:2020rjk} and in-medium potential models, see, \eg, Refs.~\cite{Rapp:2008tf,Mocsy:2013syh} for reviews. Most studies to date have focused on using point meson operators, \ie, meson operators in which  the quark and the antiquark fields are created at the same spatial point,  which give rise to the standard spectral functions. However, these correlators have overlap with all possible states containing the $Q\bar Q$ pair, including states with large invariant mass, and turned out to be not particularly sensitive to the medium modifications of quarkonium properties~\cite{Rapp:2008tf,Mocsy:2013syh}.
To enhance the sensitivity to the bound-state regime, recent lQCD studies~\cite{Larsen:2019bwy,Larsen:2019zqv} have 
utilized correlators with extended meson operators that have a larger overlap with the bottomonium state of interest but a reduced one with high-lying states. As a result, these correlators are more sensitive to medium effects at the scale of the temperature than those using point sources.

A quantitative extraction of in-medium quarkonium properties from lQCD data is not always straightforward and can result in significant sensitivity on the extraction method~\cite{Larsen:2019bwy}. Thus, a better control over the functional dependencies is needed, especially with guidance from underlying microphysics. Toward this, in the present study we will analyze lQCD results for bottomonium correlators with extended operators employing the thermodynamic $T$-matrix as a quantum many-body approach to describe the strongly coupled QGP (sQGP).  
It is based on a Hamiltonian with a non-perturbative 2-body color potential as input~\cite{Megias:2005ve} and selfconsistently evaluates the 1- and 2-body correlation functions~\cite{Mannarelli:2005pz,Riek:2010py,Liu:2017qah,Tang:2023lcn}. 
In the color-singlet channel, the interaction kernel reduces to the Cornell potential in vacuum, while its finite-temperature modifications have been constrained by lQCD data for the HQ free energy~\cite{Liu:2017qah}, quarkonium correlators~\cite{Riek:2010py}, the equation of state (EoS)~\cite{Liu:2017qah}, and most recently by static Wilson line correlators (WLCs)~\cite{Tang:2023tkm}. In particular, the selfconsistent solution of one- and two-body correlation functions under full consideration of the emerging broad spectral functions does not impose a limitation on the pertinent collision widths which is an important prerequisite for describing a strongly coupled quantum system. The additional constraints from lQCD on ground and excited bottomonium states through correlators with extended operators provide a further test of the suitability of the $T$-matrix approach to sQGP structure while also improving the precision for the resulting quarkonium spectral functions and HQ transport properties. 

The remainder of this article is organized as follows. In Sec.~\ref{sec_TM} we revisit the essential elements of the thermodynamic $T$-matrix approach  to the sQGP relevant to our study (Sec.~\ref{ssec_basic}) and derive the pertinent expression of the correlation functions with extended quark sources formulated in momentum space (Sec.~\ref{ssec_ext-op}). In Sec.~\ref{sec_vac}, we introduce a slight amendment of the vacuum potential that allows us to improve the vacuum spectroscopy of bottomonia and thus the accuracy of the in-medium applications. The main part of this paper is contained in Sec.~\ref{sec_fits} where we carry out selfconsistent $T$-matrix calculations by varying the underlying in-medium potential to achieve a combined fit to lQCD data for the equation of state (Sec.~\ref{ssec_EOS}), static Wilson line correlators (Sec.~\ref{ssec_WLC}) and bottomonium correlators with extended operators (Sec.~\ref{ssec_extended}). In Sec.~\ref{sec_transport} we quantify the update of the in-medium potential for our calculation of the charm-quark transport coefficients and compare these with recent lQCD results. Our summary and conclusions are given in Sec.~\ref{sec_concl}.

\section{$T$-matrix and Spectral Functions with Extended Operators}
\label{sec_TM} 
In this section, we first give a brief review of the thermodynamic $T$-matrix approach to the sQGP (Sec.~\ref{ssec_basic}) and then derive the expression for the extended operators in this framework (Sec.~\ref{ssec_ext-op}).

\subsection{Basic Elements}
\label{ssec_basic}
In the $T$-matrix formalism one evaluates in-medium 1- and 2-body correlation functions in a quantum many-body system by resumming an infinite series of ladder diagrams, which is a prerequiuste for studying both bound and scattering states in strongly interacting media. Initially formulated for analyzing HF particles within the QGP~\cite{Mannarelli:2005pz,Cabrera:2006wh,Riek:2010fk}, this method was later expanded to systematically incorporate the light-parton sector~\cite{Liu:2017qah}, which opened the possibility to embed the HF calculations in an interacting medium that is based on the same underlying forces and can be constrained by the equation of state from lattice QCD. The method involves reducing the 4-dimensional (4D) Bethe-Salpeter equation into a simpler 3D Lippmann-Schwinger equation~\cite{Brockmann:1996xy}, which can be further reduced through a partial-wave expansion to a 1D scattering equation,
\begin{eqnarray}
\ T_{ij}^{L,a} ( z,p,p')&=&V_{ij}^{{L,a}} (p,p') +\frac{2}{\pi } \int_{-\infty}^{\infty}k^{2}dk V_{ij}^{{L,a}} (p,k)\nonumber \\
&&\times G_{ij}^{0} (z,k)T_{ij}^{L,a} ( z,k,p') \ .
\label{eq_Tmat}
\end{eqnarray}
The intermediate 2-body propagator is given by 
\begin{eqnarray}
G_{i j}^{0}(z, k)&=& \int_{-\infty}^{\infty} d \omega_{1} d \omega_{2} \frac{\left[1 \pm n_{i}\left(\omega_{1}\right) \pm n_{j}\left(\omega_{2}\right)\right]}{z-\omega_{1}-\omega_{2}} \nonumber \\
&&\times\rho_{i}\left(\omega_{1},k\right) \rho_{j}\left(\omega_{2}, k\right) \ ,
\label{G2}
\end{eqnarray}
with single-particle spectral functions
\begin{equation} 
\rho_{i}\left(\omega, k\right)=-\frac{1}{\pi} \operatorname{Im}G_{i}(\omega+i \epsilon,k) \ ,
\label{rhoi}
\end{equation}
and single-particle propagators 
\begin{equation}
G_{i}(\omega,k) =  1/[\omega-\varepsilon_i(k) -\Sigma_i(\omega,k)]  \ .
\label{Gi}
\end{equation}
In Eq.~(\ref{eq_Tmat}), the potential $V_{ij}^{L,a}$ between partons $i$ and $j$ is specified by the angular momentum $L$ and color channel $a$. The functions $n_{i}$ represent Bose ($+$) or Fermi ($-$) distributions for gluons or anti/quarks, respectively. The on-shell energies are denoted as $\varepsilon_i = \sqrt{M_{i}^2 + k^2}$, where the particle mass $M_{i}$ is given by $M_{i} = -\frac{1}{2} \int \frac{d^3 \mathbf{p}}{(2\pi)^3} V_{i \bar{i}}^{a=1}(\mathbf{p}) + M^0_{i}$, which includes a bare mass, $M^0_{i}$, and a selfenergy component from the color-singlet ($a=1$) potential, known as the ``Fock term". The variables $p$ and $p'$ denote the magnitudes of the initial and final momenta in the center-of-mass (c.m.) frame. The in-medium single-parton selfenergies, $\Sigma_i(\omega,k)$, are computed by closing the $T$-matrix with a one-parton propagator in the medium. The 1D integral equation (\ref{eq_Tmat}) can be solved by discretizing its 3-momenta followed by a matrix inversion~\cite{Liu:2017qah}.

For the static in-medium potential in the color-singlet channel we make the ansatz~\cite{Megias:2005ve}
\begin{equation}
\widetilde{V}(r,T) = -\frac{4}{3} \alpha_{s} [\frac{e^{-m_{d} r}}{r} + m_{d}] -\frac{\sigma}{m_s} [e^{-m_{s} r-\left(c_{b} m_{s} r\right)^{2}}-1] \ ,
\label{eq_Vr}
\end{equation}
which in vacuum reduces to the well-known Cornell potential, $\widetilde{V}(r) = -\frac{4}{3} \frac{\alpha_{s}}{r}  +\sigma r$, with the strong coupling constant $\alpha_s$ and string tension $\sigma$.
The parameters $m_d$ and $m_s$ are the Debye screening masses for the short-range color-Coulomb and long-range string interactions, respectively. The parameter $c_b$ is introduced in the exponential's quadratic term of the confining potential to mimic in-medium string breaking at large distances while ensuring a smooth functional form suitable for numerical implementation~\cite{Liu:2017qah}. 
The potential is subtracted by its infinite-distance value, $V(r,T)=\tilde{V}(r,T) -\tilde{V}(\infty,T)$, and converted into momentum space through a Fourier transform. 
The potentials involving partons with finite masses acquire relativistic corrections obtained from the Lorentz structure of the interaction vertex~\cite{Riek:2010fk},
\begin{equation}
\begin{aligned}
&V_{ij}\left(\mathbf{p}, \mathbf{p}^{\prime}\right)=\mathcal{R}_{ij}^{vec} V^{vec}\left(\mathbf{p}-\mathbf{p}^{\prime}\right)+\mathcal{R}_{ij}^{sca}V^{sca}\left(\mathbf{p}-\mathbf{p}^{\prime}\right) \ .
\end{aligned}
\label{eq_V}
\end{equation}
For the static vector ($V^{vec}$) and scalar ($V^{sca}$) potentials one has
\bea
\mathcal{R}_{ij}^{vec}&=&\sqrt{1+\frac{p^{2}}{\varepsilon _{i}(p)\varepsilon_{j}(p)}}\sqrt{1+\frac{p'^{2}}{\varepsilon _{i}(p')\varepsilon _{j}(p')}} \nonumber ,
\\
\mathcal{R}_{ij}^{sca}&=&\sqrt{\frac{M_{i}M_{j}}{\varepsilon_i(p)\varepsilon_{j}(p)}}\sqrt{\frac{M_{i}M_{j}}{\varepsilon_i(p')\varepsilon_{j}(p')}} \ .
\label{eq_R correction}
\eea
Following Ref.~\cite{Tang:2023lcn}, we allow for a scalar-vector mixing (characterized by a mixing coefficient, $\chi$) in the confining potential, \ie, $V^{vec}=V_\mathcal{C}+(1-\chi)V_\mathcal{S}$ and $V^{sca}=\chi V_\mathcal{S}$, inspired by studies in Refs.~\cite{Szczepaniak:1996tk,Brambilla:1997kz,Ebert:2002pp}. 
For $\chi=1$, the confining potential is purely scalar, which is a common assumption (along with a purely vector Coulomb potential)~\cite{Mur:1992xv,Lucha:1991vn}, while for $\chi<1$ one has a vector admixture.
In Ref.~\cite{Tang:2023lcn}, we have found that, by comparing the results for $\chi$=1 and $\chi$=0.6, that the latter improves the spin-induced splittings in vacuum charmonium and bottomonium spectroscopy
appreciably~\cite{ParticleDataGroup:2018ovx}. In our later studies of static WLCs~\cite{Tang:2023tkm}, a larger mixing coefficient, $\chi$=0.8, has been used (this was constrained by the fact that for the large in-medium potentials favored by the WLCs gluon condensation could occur which currently is beyond the scope of our selfconsistent $T$-matrix calculations). However, for $\chi$=0.8 one still obtains a marked improvement in the hyperfine splittings in the vacuum spectroscopy. 
In addition, the resulting prediction for the HQ diffusion coefficient is in better agreement with lQCD data~\cite{Altenkort:2023oms,Altenkort:2023eav} than for $\chi=1$~\cite{Tang:2023lcn,Tang:2023tkm}. Therefore, we continue to use $\chi$=0.8 in the present study.
The potential is then implemented into different color channels with pertinent Casimir coefficients~\cite{Liu:2017qah,Tang:2023lcn}.

\subsection{Bottomonium Correlators with Extended Operators}
\label{ssec_ext-op} 
Bottomonium correlators in lQCD have recently been studied with so-called ``extended operators" (in Coulomb gauge)~\cite{Larsen:2019zqv} which amount to choosing the sources of the $b$ and $\bar{b}$ quark at a finite spatial separation, $r$, 
\begin{equation}
O_i(\mathbf{x}, \tau)=\sum_{\mathbf{r}} \Psi_i(\mathbf{r}) \bar{q}(\mathbf{x}+\mathbf{r}, \tau) \Gamma q(\mathbf{x}, \tau),
\label{Oext}
\end{equation}
where $q(\mathbf{x}, \tau)$ or $\bar{q}(\mathbf{x}, \tau)$ stands for the quark or antiquark field at position $x$ and Euclidean time $\tau$, and
$\Gamma= 1,\gamma^\mu$ are the vertex operators for the mesonic scalar and vector channels, respectively. The wave functions $\Psi_i$ in 
Eq.~(\ref{Oext}) are chosen to provide an overlap that is tailored to the bottomonium  state of interest ($i$ = $1S$, $2S$, $3S$, $1P$, $2P$).     
In practice, wave functions obtained from solving the Schr\"odinger equation with a vacuum Cornell potential and a
nominal value of the bottom-quark mass are employed.
The potential and the bottom-quark mass for this purpose are generally different from the one used in the $T$-matrix calculations, since $\Psi_i$ is merely a trial wave
function and the physics information in the spectral function does not depend on these parameters.
In the lattice study of Ref.~\cite{Larsen:2019zqv} the strong coupling constant, string tension and bottom-quark mass 
used to obtain $\Psi_i$ are chosen as $\alpha_s=0.24$, $\sigma=(468~\textup{MeV})^2$ and $M_b=4.676$~GeV, respectively. 

The bottomonium correlators, defined as 
\beq
G_{i j}(\tau)=\left\langle O_i(\mathbf{x},\tau) O_j^{\dagger}(\mathbf{0},0)\right\rangle \ ,
\eeq
characterize the propagation of a $b\bar{b}$ pair from $(\mathbf{0},0)$ to $(\mathbf{x}, \tau)$. To suppress the mixing with other states, optimized operators have been introduced in Ref.~\cite{Larsen:2019zqv}, 
\beq
G_{\alpha}(\tau)=\left\langle \tilde{O}_\alpha(\mathbf{x},\tau) \tilde{O}_\alpha^{\dagger}(\mathbf{0},0)\right\rangle \ ,
\eeq
as a linear combination of the original ones, $\tilde{O}_\alpha=\Omega_{\alpha j}O_j$, such that $\left\langle\tilde{O}_\alpha(\mathbf{x},\tau) 
\tilde{O}_\beta^{\dagger}(\mathbf{x},0)\right\rangle \propto \delta_{\alpha, \beta}$. 
The matrices $\Omega_{\alpha j}$ were obtained by solving the generalized eigenvalue problem given by 
\beq
G_{i j}(\tau) \Omega_{\alpha j}=\lambda_\alpha\left(\tau, \tau_0\right) G_{i j}\left(\tau_0\right) \Omega_{\alpha j} \ .
\eeq

The quarkonium spectral function, $\rho_{Q\bar{Q}}^{\alpha}(E,T)$, is related to the Euclidean time correlation function through a Laplace transform,
\begin{equation}
G_\alpha(\tau, T)=\int_{-\infty}^{\infty} d E \rho_{Q\bar{Q}}^{\alpha}(E, T) e^{-E \tau} \  ,
\end{equation}
where $E$ is the total energy of quarkonium in the c.m. frame.
In Ref.~\cite{Larsen:2019zqv}, the temperature-independent high-energy contribution (continuum) to the spectral function has been subtracted since that part is not related to the in-medium bound state properties. 
The (subtracted) quarkonium spectral function in the $T$-matrix approach is given by the imaginary part of correlation function in energy representation,  
\begin{equation}
\rho_{Q\bar{Q}}^{\alpha}(E, T)= - \frac{1}{\pi}\operatorname{Im}G_\alpha(E+i\epsilon,T) \ .
\end{equation}
The correlation function is composed of a free and an interacting part: $G_\alpha=G_\alpha^0+\Delta G_\alpha$. For the case of extended operators, one can show that they are given by (see Appendix~\ref{sec_sfex})
\begin{eqnarray}
G_\alpha^0(E)&= &N_{f}N_{c}\int \frac{d^3p}{(2\pi )^3}\mathcal{R}_{Q\bar{Q}}^{sca} a_{free}(p) G_{Q\bar{Q}}^0(E,p){\Psi}^2(p) 
\nonumber \\
\Delta G_\alpha(E)&= &\frac{N_f N_c}{\pi ^3}\int  dp p^2 \mathcal{R}_{Q\bar{Q}}^{sca} G_{Q\bar{Q}}^0(E,p)\nonumber \\
&&\times\int dp' p'^2 G_{Q\bar{Q}}^0(E,p'){\Psi}(p){\Psi}(p')\nonumber \\
&&\times[a_0(p,p')T_{Q\bar{Q}}^{0}+a_1(p,p')T_{Q\bar{Q}}^{1}] \ ,
\label{eq_Cmed}
\end{eqnarray}
where $T_{Q\bar{Q}}^{0,1}$ are the $T$-matrices in $S$- and $P$-wave channels, respectively, and ${\Psi}(\mathbf{p})=\int d^{3}\mathbf{r}e^{-i\mathbf{p}\cdot \mathbf{r}}\Psi(\mathbf{r})$ is the wave function for the extended operator in momentum space.
The leading-order in $1/M_b$ of the $a_{0,1,free}$ coefficients for different mesonic channels are summarized in Tab.~\textup{V} of Ref.~\cite{Tang:2023lcn}.

The $T$-matrix formalism provides a selfconsistent solution of the 1- and 2-parton correlation functions. The inelastic reaction channels are included via the selfenergies of the individual HQ propagators, whose absorptive parts underlie the width of the bound states. However, for deeply bound states interference effects  for inelastic reactions have been found to be quantitatively important. In essence, the amplitudes for absorption on quark and antiquark within the
bound state interfere leading to a suppression of the total width (this phenomenon is sometimes also referred to as an ``imaginary part" of the potential~{\cite{Laine:2006ns}, or more generically a wave function effect in the reaction rate~\cite{Bhanot:1979vb,Song:2005yd}).
Specifically, in the color-singlet channel, a compact 
$Q\bar{Q}$ state effectively becomes colorless, thereby reducing interactions with the colored medium partons.
In the $T$-matrix formalism these effects correspond to 3-body diagrams which are a priori not included. However, in Ref.~\cite{Liu:2017qah} an effective implementation has been introduced via a complex potential, 
\begin{equation}
V(r)\rightarrow V(r)+\Sigma_{Q \bar{Q}}\phi(r) \  ,
\label{eq_Vcomplex}
\end{equation}
where the non-interacting two-body selfenergy, $\Sigma_{Q\bar{Q}}$, has been added in connection with the  ``interference function", $\phi(r)$. 
The former is related to the two-body propagator by
\begin{equation}
\left[G_{Q \bar{Q}}^0(E)\right]^{-1}=E-\widetilde{V}(r\rightarrow\infty)-\Sigma_{Q \bar{Q}}(E) \ ,
\label{eq_Sigma_QQ}
\end{equation}
while $\phi(r)$ amounts to an $r$-dependent suppression factor which
vanishes at $r\to0$ and tends to 1 at $r\to \infty$; its functional form is adopted from the perturbative calculations in Ref.~\cite{Laine:2006ns} but with an extra scale factor to mimic nonperturbative effects.
Equation~(\ref{eq_Vcomplex}) is then Fourier-transformed to momentum space,
\begin{equation}
V(\mathbf{p}-\mathbf{p'})\rightarrow V(\mathbf{p}-\mathbf{p'})+\Sigma_{Q \bar{Q}}(E,p)\tilde{\phi}(\mathbf{p}-\mathbf{p'}),
\end{equation}
and serves as the input to the $T$-matrix in Eq.~(\ref{eq_Tmat}), where $\tilde{\phi}(\mathbf{p}-\mathbf{p'})$ denotes the Fourier transform of $\phi(r)$.
As discussed in Ref.~\cite{Liu:2017qah}, the $T$-matrix with interference effects is still analytic but no longer positive-definite. Nevertheless, the quarkonium correlators and spectral functions remain positive definite.

From the correlators with extended operators  one can define an effective mass via~\cite{Larsen:2019zqv}
\begin{equation}
a M_{\mathrm{eff}}(\tau, T)=\ln \left(G_\alpha(\tau, T) / G_\alpha(\tau+a, T)\right),
\label{eq_Meff}
\end{equation}
with $a$ the lattice spacing. 
In the limit $a \rightarrow 0$ the effective mass can be written as $M_{\mathrm{eff}}(\tau, T)=\partial_{\tau} G_\alpha(\tau, T)$. For a spectral function $\rho_{Q \bar Q}^{\alpha}$ that is dominated by a single narrow peak 
the effective mass for very small $\tau$, 
$ M_{\mathrm{eff}}(\tau\to0, T)$, is closely related to the mass of the bound state, while the slope of $M_{\mathrm{eff}}(\tau, T)$ is related to the width of the peak in $\rho_{Q\bar{Q}}^{\alpha}$.


\section{Vacuum Potential Constrained by Bottomonium Spectroscopy}
\label{sec_vac} 
In previous works our starting point has been a vacuum potential that accurately describes lQCD results for the $Q{\bar Q}$ free energy. However, one might argue that in the presence of a nontrivial vacuum structure this may not necessarily be a requirement as vacuum polarization diagrams may be present in the free energy, and thus the driving kernel of the scattering equation may not be the same quantity. 
In addition, close to the open HF threshold, hadronic interactions are expected to play a role for the masses of the weakly bound states that are not included in our quark-based description.
Since we do not compute vacuum and hadron structure effects selfconsistently in our current formalism, we have explored whether variations in the potential can lead to any improvement in the predictions for vacuum spectroscopy. 

As in Refs.~\cite{Tang:2023lcn}, the quarkonium masses are extracted from the pole positions of spectral functions (imaginary parts of the correlation functions) with point operators which are given by the correlation functions in Eq.~(\ref{eq_Cmed}) but without the wave function ${\Psi}$. 
With  $\alpha_s=0.31$ and $\sigma=0.25$\,GeV$^2$ (compared to 0.27 and 0.225\,GeV$^2$, respectively, in our previous work) we indeed find an appreciable improvement in describing the vacuum masses of charmonia and bottomonia as listed by the Particle Data Group (PDG)~\cite{ParticleDataGroup:2018ovx}, while keeping the mixing parameter at $\chi$=0.8.
The masses of almost all states below the open-bottom threshold are now within 20\,MeV of the experimental values (with the exception of the $\Upsilon(2S)$), see Table~\ref{tab_Ymass} for the comparison of vacuum bottomonium spectroscopy between $T$-matrix results and experimental data. 
The new potential slightly deviates from the vacuum free energy obtained from lQCD at large distances (and from our previous version), see  Fig.~\ref{fig_vac-pot}.
The vacuum charmonium masses show a similar improvement relative to the previous potential.
\begin{table}[!t]
\setlength{\tabcolsep}{3pt}
\begin{center}
\def\temptablewidth{0.8\textwidth}
\begin{tabular}{c c c c c c}
\hline
\hline
 Channel  & Particle  & Expt. & \begin{tabular}[c]
{@{}l@{}}Theoretical \\
\\ 
$\chi=0.8$\\ 
$M_b^0=4.676$\\ 
$M_b=5.316$\end{tabular} & 
\begin{tabular}[c]
{@{}l@{}}Theoretical \\(previous)~\cite{Tang:2023tkm}\\ 
$\chi=0.8$\\ 
$M_b^0=4.681$\\ 
$M_b=5.257$\end{tabular} & 
 \\ \hline
\multirow{2}{*}     {S}        & $\chi_{b0} (1P)$   & 9.859    & 9.862      & 9.870           \\
                               & $\chi_{b0} (2P)$   & 10.233   & 10.250     & 10.221           \\
                     PS        &$\eta_b(1S)$        & 9.399    & 9.417      & 9.474          \\
 \multirow{3}{*}    {V}        & $\Upsilon (1S)$    & 9.460    & 9.460      & 9.517            \\
                               & $\Upsilon (2S)$    & 10.023   & 9.994      & 9.997             \\
                               & $\Upsilon (3S)$    & 10.355   & 10.370     & 10.339             \\
                   AV$_1$      & $h_b (1P)$         & 9.899    & 9.890      & 9.895             \\
 \multirow{3}{*}  {AV$_2$}     & $\chi_{b1} (1P)$   & 9.893    & 9.887      & 9.888             \\
                               & $\chi_{b1} (2P)$   & 10.255   & 10.272     & 10.247            \\
\multirow{2}{*}{T}             & $\chi_{b2} (1P)$   & 9.912    & 9.900      & 9.898             \\
                               & $\chi_{b2} (2P)$   & 10.269   & 10.287     & 10.250          \\  
\hline
\hline
\end{tabular}
\end{center}
\caption{$T$-matrix results for the masses (in GeV) of bottomonium states below the open-bottom threshold, compared to their experimental values~\cite{ParticleDataGroup:2018ovx} (third column) corresponding to the two Cornell potentials shown in Fig.~\ref{fig_vac-pot}. Spin-spin and spin-orbit interactions are included, and the confining force has a 20\% admixture of a Lorentz-vector structure (corresponding to a mixing coefficient of $\chi$=0.8); $M_b^0$ and $M_b$ denote the bare and constituent bottom-quark masses, respectively. The $\chi^2$ value improves by about 40\% in the updated fit.}
\label{tab_Ymass}
\end{table}

\begin{figure}[tbp]
\begin{minipage}[b]{1.0\linewidth}
\centering
\includegraphics[width=0.85\textwidth]{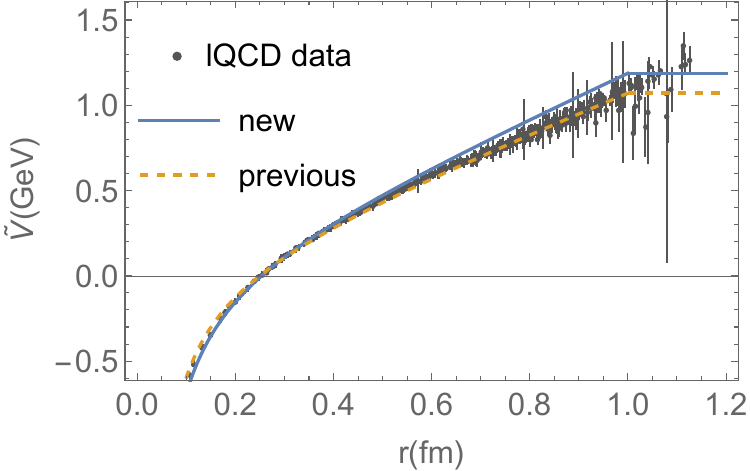}
\end{minipage}
\caption{The vacuum potentials used in our current (solid blue line) and previous (orange dashed line)~\cite{Tang:2023tkm} study as a function of distance, in comparison to vacuum free-energy data from lattice QCD~\cite{HotQCD:2014kol}.} 
\label{fig_vac-pot}
\end{figure}

\section{In-Medium Potentials Constrained by lQCD}
\label{sec_fits} 
We now turn to analyzing the lQCD results for the bottomonium correlators with extended operators. As mentioned above, key inputs are the potential as well as the interference functions. Since we want to maintain the consistency with our previous fits to WLCs, and since the results are embedded into the light-parton sector to maintain a realistic equation of state, our procedure is based on a variational method that involves two iterative loops, as follows:
Initially, we refine the in-medium potential using the effective mass ($\Meff$) of bottomonum correlators defined in Eq.~(\ref{eq_Meff}) as well as the effective mass of the WLCs ($m_1$), which will be discussed in Sec.~\ref{ssec_WLC}. These quantities, derived from the $T$-matrix approach, are fitted to lQCD data. Using the adjusted potentials, we proceed to calculate the QGP EoS, modifying the main parameters in the light-parton sector, which are the ``bare" masses of the in-medium light quarks and gluons. The computation of the EoS involves both one-body spectral functions and two-body scattering amplitudes, requiring the solution of a selfconsistency problem carried out through numerical iteration.
Following this, we recompute the HQ self-energies from the heavy-light $T$-matrices and reintegrate these results into the calculations of $\Meff$ and $m_1$ using the in-medium quarkonium spectral functions. We continue refining the in-medium potential by refitting the effective masses obtained in the $T$-matrix approach to the corresponding lQCD data. The adjusted potential is then employed to fit the EoS again. This loop is repeated until convergence is achieved.

In the remainder of this section, we discuss our numerical fits of the EoS in Sec.~\ref{ssec_EOS} and to the WLCs in Sec.~\ref{ssec_WLC}, followed by a discussion of the results for the bottomonium correlators with extended operators and a comparison of their spectral functions to those with point operators in Sec.~\ref{ssec_extended}.

\subsection{Equation of State}
\label{ssec_EOS}
The equation of state is encoded in the pressure, $P(T,\mu)$, of a many-body system as a function of temperature and chemical potential; it is connected to the thermodynamic potential per unit volume via $\Omega=-P$. For interacting quantum systems, the pressure can be computed diagrammatically within the Luttinger-Ward-Baym (LWB) formalism~\cite{Luttinger:1960ua,Baym:1961zz,Baym:1962sx}. This includes an interaction contribution represented by the Luttinger-Ward functional (LWF), which constitutes a thermodynamically consistent formalism when combined with the latter resummation in the $T$-matrix. For strongly interacting systems, it is important to resum the fully dressed skeleton  diagrams in the LWF, which can be achieved with a matrix-logarithm technique~\cite{Liu:2016ysz}. This enables to account for the contributions from dynamically formed bound states and/or resonances to the EoS. The thermodynamic potential is given by~\cite{Liu:2016ysz,Liu:2017qah}
\begin{equation}
\begin{aligned}
\Omega=& \sum_{j} \mp d_{j} \int d \tilde{p}~\Big\{ \ln \left(-G_{j}(\tilde{p})^{-1}\right)\\
&+\left[\Sigma_{j}(\tilde{p})-\frac{1}{2} \log \Sigma_{j}(\tilde{p})\right] G_{j}(\tilde{p})\Big\} \ ,
\end{aligned}
\label{eq_EOS}
\end{equation}
where $\int d \tilde{p} \equiv-\beta^{-1} \sum_{n} \int d^{3} \mathbf{p} /(2 \pi)^{3}$ and $\tilde{p} \equiv\left(i \omega_{n} ,\mathbf{p}\right)$ with $\beta=1/T$. The summation in Eq.~(\ref{eq_EOS}) includes all light-parton channels characterized by the spin-color degeneracy $d_j$, with the $\mp$ sign distinguishing bosons ($-$) from fermions ($+$). The three components of Eq.~(\ref{eq_EOS}) $ - $ $\ln (-G^{-1})$, $\Sigma G$, and $\log \Sigma G$ $ - $ correspond to the contributions from quasiparticles, their selfenergies, and two-body interactions (LWF), respectively.

The final converged results are very similar to our previous work in Ref.~\cite{Tang:2023tkm}: the contribution from the two-body interactions become more significant as the temperature decreases, driven by the formation of bound states in the attractive color channels (singlet and anti-triplet). This indicates a transition in the degrees of freedom within the system from partons to mesons and diquarks at lower temperatures, which is in agreement with the insights of earlier work~\cite{Liu:2017qah}.

\subsection{Static Wilson Line Correlators}
\label{ssec_WLC}
Next, we turn to the static WLCs in Euclidean time, which are related to the static $Q\bar{Q}$ spectral functions, $\rho_{Q\bar{Q}}$, via a Laplace transform,
\begin{equation}
W\left (r,\tau,T  \right )=\int_{-\infty}^{\infty}dE e^{-E \tau}\rho_{Q\bar{Q}}\left ( E,r,T \right) \ ,
\label{wlc}
\end{equation}
where $r$ is the separation between $Q$ and $\bar{Q}$ and $E$ their total energy (subtracted by twice bare HQ mass, numerically taken as $2M_Q^0=2\times 10^4$ GeV). The constituent static HQ mass is the sum of the bare mass and the mass shift originating from the self-energy, \ie, $M_Q=M_Q^0+\widetilde{V}(r\rightarrow\infty)/2$~\cite{Liu:2017qah} 
(the Fock term, $-\frac{1}{2} \int \frac{d^3 \mathbf{p}}{(2\pi)^3} V_{i \bar{i}}^{a=1}(\mathbf{p})$, defined below Eq.~(\ref{Gi}) reduces to $\widetilde{V}(r\rightarrow\infty)/2$ in the static limit). 
In the $T$-matrix formalism, the $Q\bar{Q}$ spectral function takes the following form~\cite{Liu:2017qah,Tang:2023tkm},
\begin{equation}
\rho_{Q\bar{Q}}\left ( E,r,T \right )=\frac{-1}{\pi}\mathrm{Im}\left [ \frac{1}{E-\widetilde{V}(r,T)-\phi(r,T)\Sigma_{Q\bar{Q}}(E,T)} \right ]. 
\label{rho_QQ}
\end{equation}
As introduced in Sec.~\ref{ssec_basic}, $\widetilde{V}$, $\Sigma_{Q\bar{Q}}$ and $\phi$ in Eq.~(\ref{rho_QQ}) are the static in-medium potential, two-body selfenergy and interference function, respectively. 

The comparison to the lattice data is performed in terms of the effective masses of the WLCs, defined as as $m_1(r, \tau, T)=-\partial_\tau \ln W(r, \tau, T)$~\cite{Bala:2021fkm}. 
As for the correlators with extended operators discussed in Sec.~\ref{ssec_ext-op}, the temperature-independent parts at high energies, stemming from excited states related to hybrid potentials, are subtracted from the lattice WLCs.
Similar to our previous study~\cite{Tang:2023tkm}, the WLCs from the $T$-matrix show a fair overall agreement with the lQCD results for both the intercept at $\tau\rightarrow 0$ and the slopes of $m_1$, although larger discrepancies are observed at the highest temperature. The main reason for this is that the stronger input potential leads to an unstable fitting procedure due to the emergence of glueball condensation, which limits our efforts to improve the fits (see Ref.~\cite{Tang:2023tkm} for more details).



\subsection{Bottomonium Correlators with Extended Operators}
\label{ssec_extended}
The final results of our fits (which encompass the EoS and WLCs displayed in the previous two sections as well) to the effective masses of bottomonium correlators with extended operators are summarized in Fig.~\ref{fig_Meff} for three temperatures for
which lattice data are available~\cite{Larsen:2019bwy}.
In performing this comparison one should keep in mind that in lattice-NRQCD  (Non-Relativistic Quantum Chromodynamics) calculations one does not obtain the bottomonium masses but energy levels of
different bottomonium states, which depend on the details of the 
lattice-NRQCD formulation. Therefore, in these calculations the energy
levels are given with respect to a baseline energy level.
In Ref.~\cite{Larsen:2019zqv} the spin-averaged energy of the 1S bottomonia, $\bar{E}_{1S} = (E{\eta_b} + 3E_\Upsilon)/4$, is used as the baseline, and
the energy levels of other bottomonium states are 
given with respect to this baseline.
Therefore, in the present work we use the spin-averaged binding energy of
$1S$ bottomonia as reference point for the effective masses.
However, this is still not a one-to-one correspondence between
the binding energies in $T$-matrix approach and the energy levels of NRQCD.
Thus we use a smaller value for the relative shift to perform
the comparison with the lattice data:
$\bar{E}_{1S}\rightarrow\bar{E}_{1S}-0.12$GeV.
A fair overall agreement with the lQCD results is achieved for both the intercept at $\tau\rightarrow 0$ and the slopes of $\Meff$. 
 \begin{figure*}[tbp]
\centering
\includegraphics[width=0.97\textwidth]{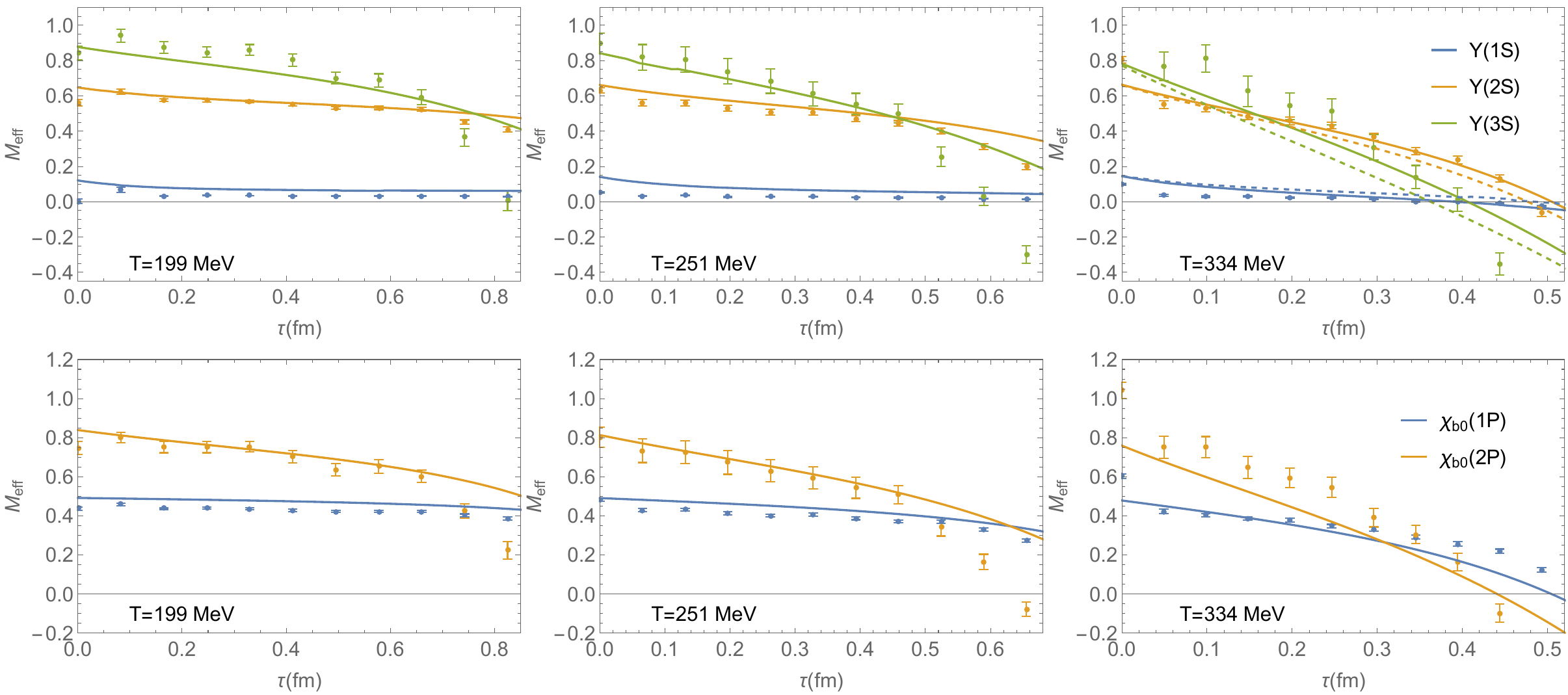}
\caption{The effective masses of bottomonium correlators with extended operators from the $T$-matrix (lines) as a function of imaginary time at different temperatures compared to the corresponding 2+1-flavor lQCD data~\cite{Larsen:2019zqv} (dots). Upper panels: the effective masses for $\Upsilon(1S)$ (blue), $\Upsilon(2S)$ (orange) and $\Upsilon(3S)$ (green) states. Lower panels: the effective masses for $\chi_{b0}(1P)$ (blue) and $\chi_{b0}(2P)$ (orange) states.  The dashed lines in the upper right panel are the results when using the old interference function (dashed lines in Fig.~\ref{fig_phi}).} 
\label{fig_Meff}
\end{figure*}

The underlying in-medium input potentials and their parameters are displayed in Fig.~\ref{fig_V}. The potentials show essentially no screening for $r<0.8$ fm, consistent with recent lQCD studies of the energy of static $Q\bar{Q}$ pairs~\cite{Bala:2021fkm,Bazavov:2023dci}.
Compared to our previous studies~\cite{Tang:2023tkm} where the potentials were only constrained by the QCD EoS and static WLCs, the potentials in this work are a little weaker at large distances (more prominent for $T$=199\,MeV) as the reduction of confining potential due to a larger $m_s$ outweighs its enhancement due to a larger $\sigma$, but more attractive at small distances due to the enhancement of the Coulomb potential with a smaller $m_d$ and larger $\alpha_s$. However, the differences between these two studies are rather modest. More significant changes are required in the interference functions, as shown in Fig.~\ref{fig_phi}. The interference effects become overall more pronounced at intermediate and large distances (smaller values of the $\phi(r)$ functions), indicating a weakening of the coupling of the color-singlet $b\bar{b}$ state to the medium partons,  and the opposite at very small distances. To explicitly demonstrate their impact on the fits, we show in the upper right panel of Fig.~\ref{fig_Meff} the results when using the previous interference function: the agreement with the lQCD data becomes noticeably worse at large distances. The impact of the change in $\phi(r)$ at small distance will become apparent in the $1S$ spectral function discussed below.

\begin{figure*}[tbp]
\begin{minipage}[b]{1.0\linewidth}
\centering
\includegraphics[width=0.97\textwidth]{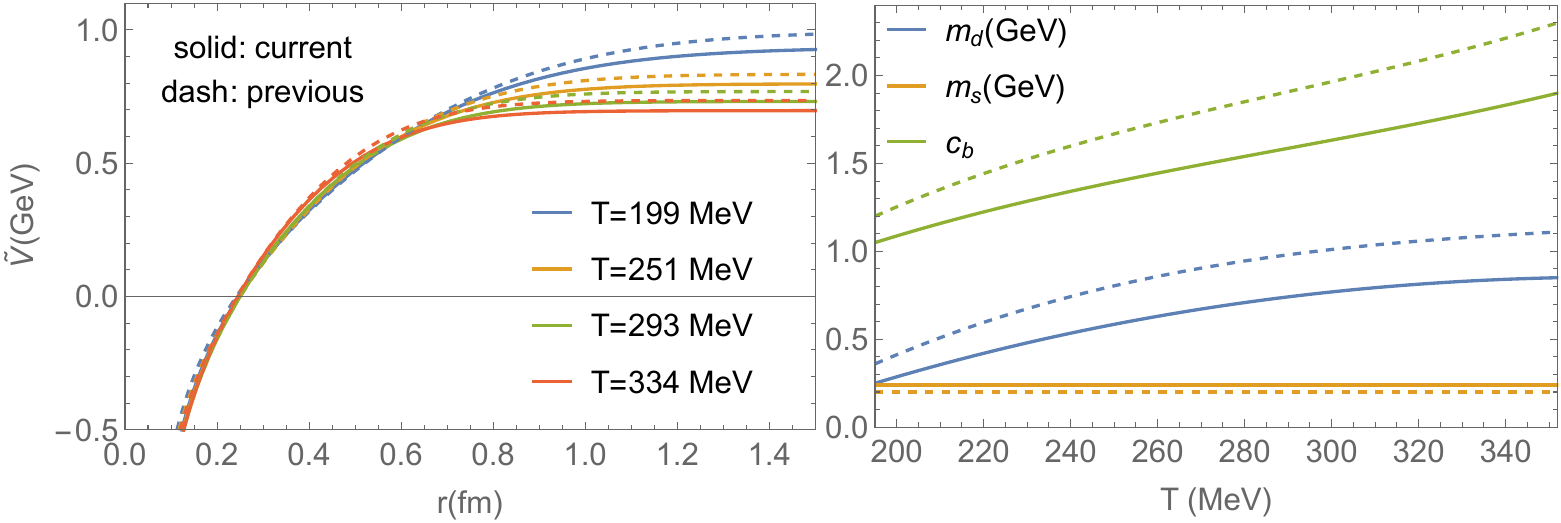}
\end{minipage}
\caption{Left panel: the in-medium potentials as a function of distance at different temperatures. Right panel: potential parameters as a function of temperature. The solid and dashed lines correspond to our current and previous studies~\cite{Tang:2023tkm}, respectively.} 
\label{fig_V}
\end{figure*}

\begin{figure}[tbp]
\centering
\includegraphics[width=0.45\textwidth]{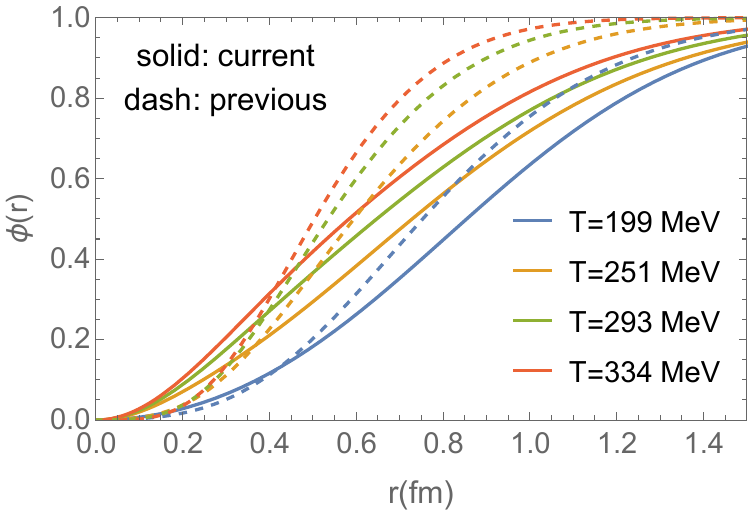}
\caption{The interference function vs.~distance at different temperatures. The solid and dashed lines correspond to our current and previous study~\cite{Tang:2023tkm}, respectively.} 
\label{fig_phi}
\end{figure}

To scrutinize and interpret the bound-state properties underlying our calculations, we compare the bottomonium spectral functions with extended and point operators (without wave functions, $\Psi$, in Eq.~(\ref{eq_Cmed}))
in Fig.~\ref{fig_SFbb}. 
For the $1S$ state, the spectral functions corresponding to both types of meson
operators agree well with each other up to the highest considered temperature of 334\,MeV.
In the case of extended operators, the $2S$ and $3S$ spectral functions show the presence of peak structures up to $T=334$ MeV. However, as the 
temperature increases these structures become
quite broad and visibly non-symmetric for higher-lying bottomonium states. 
In fact, the width of these peaks is comparable or even larger than 
the mass difference of various $\Upsilon(nS)$ bottomonium states. Therefore, the interpretation of these peaks in terms of bound states is questionable. 
The $S$-wave spectral functions corresponding to 
point meson operators convey a different message. The $\Upsilon(2S)$ state can be clearly seen at $T$=199\,MeV, but is difficult to identifyat higher temperature. Furthermore, no structures that can be associated with $\Upsilon(3S)$ 
can be identified in the spectral function corresponding to the point meson operators. This may imply that the $\Upsilon(2S)$ melts at temperatures between 200 MeV and 250 MeV.

\begin{figure*}[tbp]
\begin{minipage}[b]{1.0\linewidth}
\centering
\includegraphics[width=0.99\textwidth]{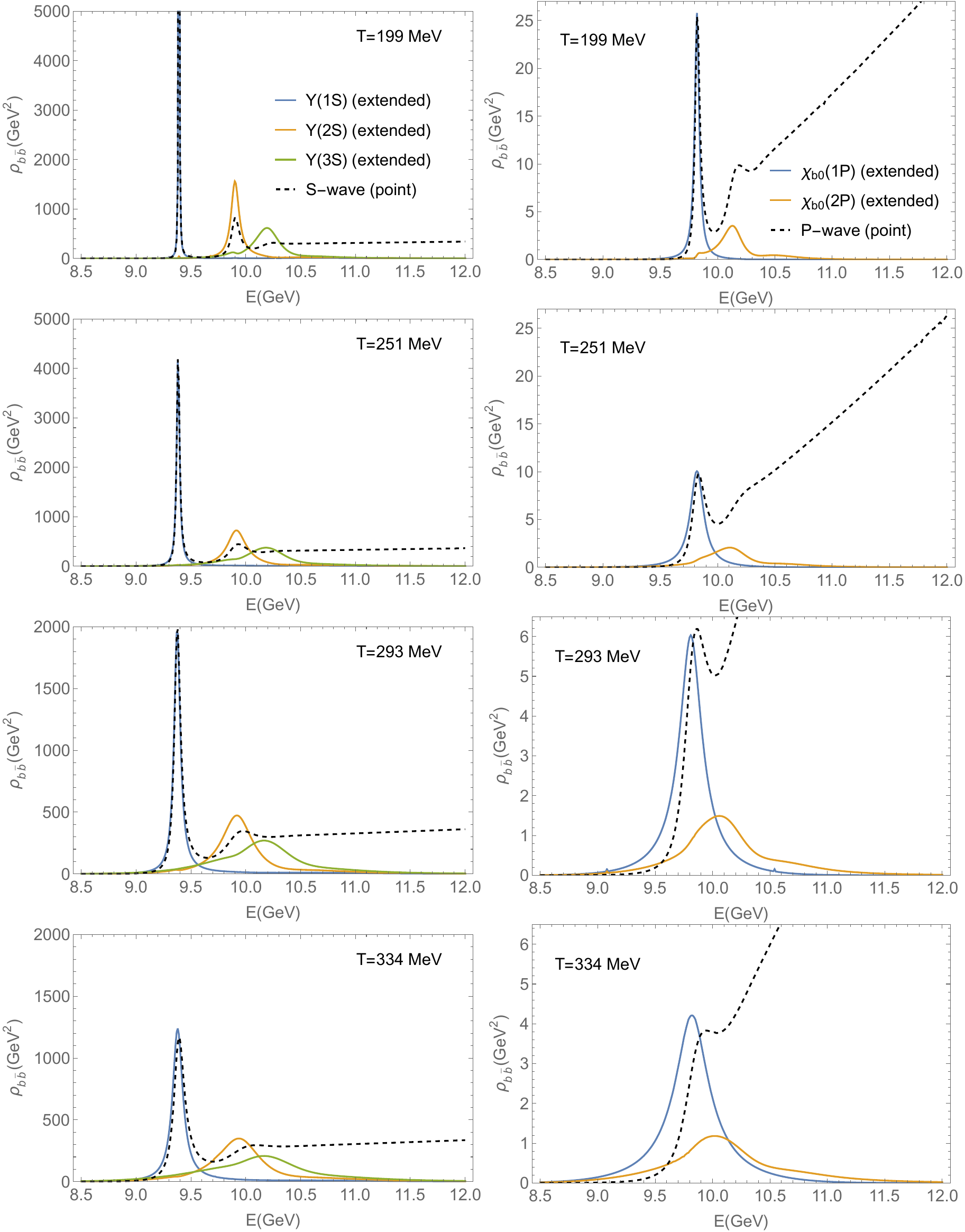}
\end{minipage}
\caption{The bottomonium spectral functions from the $T$-matrix with extended operators as a function of total energy. Left panel: $S$-wave states $\Upsilon(1S)$, $\Upsilon(2S)$ and $\Upsilon(3S)$ (blue, orange and green lines, respectively); right panel: $P$-wave states $\chi_{b0}(1P)$ (blue lines) and $\chi_{b0}(2P)$ (orange lines). 
The spectral functions with point operators are shown by black dashed lines, scaled up by suitable factors to render them comparable to those with extended operators.} 
\label{fig_SFbb}
\end{figure*}

The $1P$ and $2P$ spectral functions corresponding to the extended meson operators show peak structures for all
temperatures up to $T=334$\,MeV. The peak corresponding to the $2P$ state is quite broad and visibly asymmetric already
at $T=195$\,MeV. As the temperature increases the width of this peak increases significantly. The peak corresponding to the
$1P$ state also broadens with increasing temperature. At the highest two temperatures the width of both $1P$ and $2P$ states is comparable or larger than the energy difference between them. In the $P$-wave spectral functions corresponding
to point operators we can identify the $1P$ state for $T=199$\,MeV and $251$\,MeV, but not at the highest
two temperatures shown in Fig.~\ref{fig_SFbb}. However, we do not see the peak corresponding to the $2P$ state at any temperature. Thus, this analysis suggests that the $1P$ bottomonium state melts around a temperature of $293$~MeV.
We note that whenever well-defined peaks exist in the spectral functions of point meson operators, the corresponding peak
position agrees with the one in the spectral function of extended meson operators. Therefore, these peak positions can 
be interpreted as in-medium bottomonium masses.

To further assess the disappearance of a bound state we resort to a method employed in Ref.~\cite{Cabrera:2006wh} which relies on identifying the poles in the $T$-matrix. 
Toward this end we write the solution of the $T$-matrix equation in operator form as 
\beq 
\mathbb{T}(E)=\left[\mathbbm{1}-\mathbb{V} \hat{\mathbb{G}}_{(2)}^0(E)\right]^{-1} \mathbb{V}
\label{eq_Tmmatrix}
\eeq
(in practice, this corresponds to discretizing the momentum integral with $\mathbb{V}_{m n} \equiv V\left(k_m, k_n\right)$, $\mathbb{T}(E)_{m n}\equiv T\left(E, k_m, k_n\right)$ and $\hat{\mathbb{G}}_{(2)}^0(E)_{m n} \equiv \frac{2 \Delta k}{\pi} k_m^2 G_{(2)}^0\left(E, k_m\right) \delta_{m n}$~\cite{Haftel:1970zz}). 
Defining $\mathcal{F}(E)=\mathbbm{1}-\mathbb{V} \hat{\mathbb{G}}_{(2)}^0(E)$, the requirement $\operatorname{det} \mathcal{F}(E)=0$ (for $E<E_{\mathrm{th}}$) indicates the presence of a bound state. This is so because a bound state is characterized by a pole of the scattering amplitude on the real energy axis below the threshold energy, $E_{\mathrm{th}}$ 
(this corresponds to identifying the zeros of the Jost function in $S$-matrix theory). We depict $\operatorname{det}\mathcal{F}(E)$ as a function of total energy at various temperatures, together with the imaginary parts of the $T$-matrices in Fig.~\ref{fig_detF}. As discussed in Sec.~\ref{ssec_ext-op}, the $T$-matrices with interference effect are not positive-definite at low momenta, thus we display them at somewhat larger momenta in Fig.~\ref{fig_detF} to show the peaks corresponding to the bound states more clearly. We have found that the peak locations in the $T$-matrix are not sensitive to this choice and in any case are only shown for guidance; the Jost function, as a determinant in momentum space,  does not depend on momentum.

In the $S$-wave channel at 199\,MeV, $\operatorname{det} \mathcal{F}(E)$ crosses zero three times, but only the first two ($1S$ and $2S$) are accompanied by identifiable peaks in the imaginary part and can be regarded as bound states, whereas the energy of the  third zero-crossing ($\sim10.35$GeV) is already above the $b\bar{b}$ threshold ($\sim10.23$GeV). 
Only for temperatures $T<163$ MeV we have a zero crossing below the $b\bar{b}$ threshold that can be identified with the
$3S$ state.
At the next higher temperature, $T$=251\,MeV, the $2S$ pole has also disappeared even though a peak with a width comparable or larger than the binding energy still shows up in the imaginary part of the $T$-matrix which is very similar to the pertinent point spectral function. 
We still have a pole corresponding to the $2S$ state at temperatures around 220 MeV, so its ``melting" temperature is slightly higher than this.
The $1S$  state, on the other hand, still persists at the highest temperature considered in this study.
In the $P$-wave channel we see a zero corresponding to $1P$  state at $T$=251\.MeV, but this zero is about to disappear at $T$=293\,MeV.
On the other hand, the zero corresponding to $2P$ state disappears close to 174\,MeV.\footnote{
In a very recent work, the pole analysis of the in-medium quarkonium $T$-matrices has been further developed by carrying it into the complex energy plane. True poles for the pertinent bound states can be identified yielding an unambiguous criterion for the survival of a state along with its in-medium mass and width. Somewhat surprisingly, the pertinent melting temperatures were found to be significantly larger than found here, and also much larger compared to previous estimates by comparing the width and binding energies in connection with energy uncertainty arguments.} 
We also illustrate the impact of the changes in the interference function by plotting the results for the older version in the lower right panel of Fig.~\ref{fig_detF}; here, the effect is particularly visible for the $1S$ state due to the changes in the small-distance region: the larger values from the new constraints on $\phi(r)$ lead to a $\sim$40\% increase in the width of the $1S$ bound-state peak.

\begin{figure*}[tbp]
\includegraphics[width=0.82\textwidth]{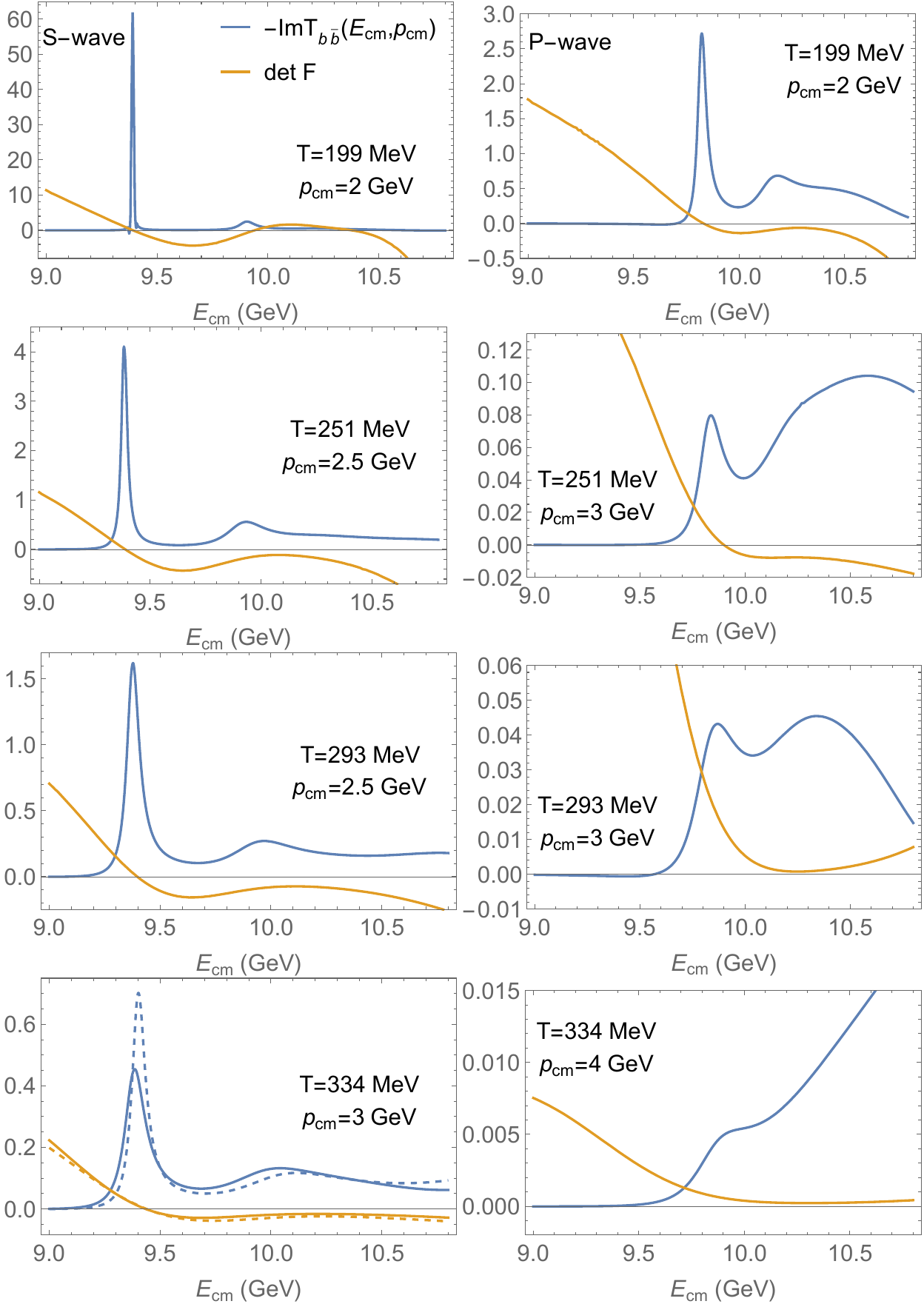}
\caption{The bottomonium $T$-matrices and determinant functions, $\operatorname{det}\mathcal{F}$, for $S$- (left) and $P$-wave (right) states as a function of c.m. energy at different temperatures. Note that the $\operatorname{det}\mathcal{F}$ functions have been scaled up by suitable factors to be comparable to the $T$-matrices. The dashed lines in the lower left panel are the results when using the old interference function (dashed lines in Fig.~\ref{fig_phi}).}
\label{fig_detF}
\end{figure*}

\begin{figure*}[!tbp]
\centering
\includegraphics[width=0.97\textwidth]{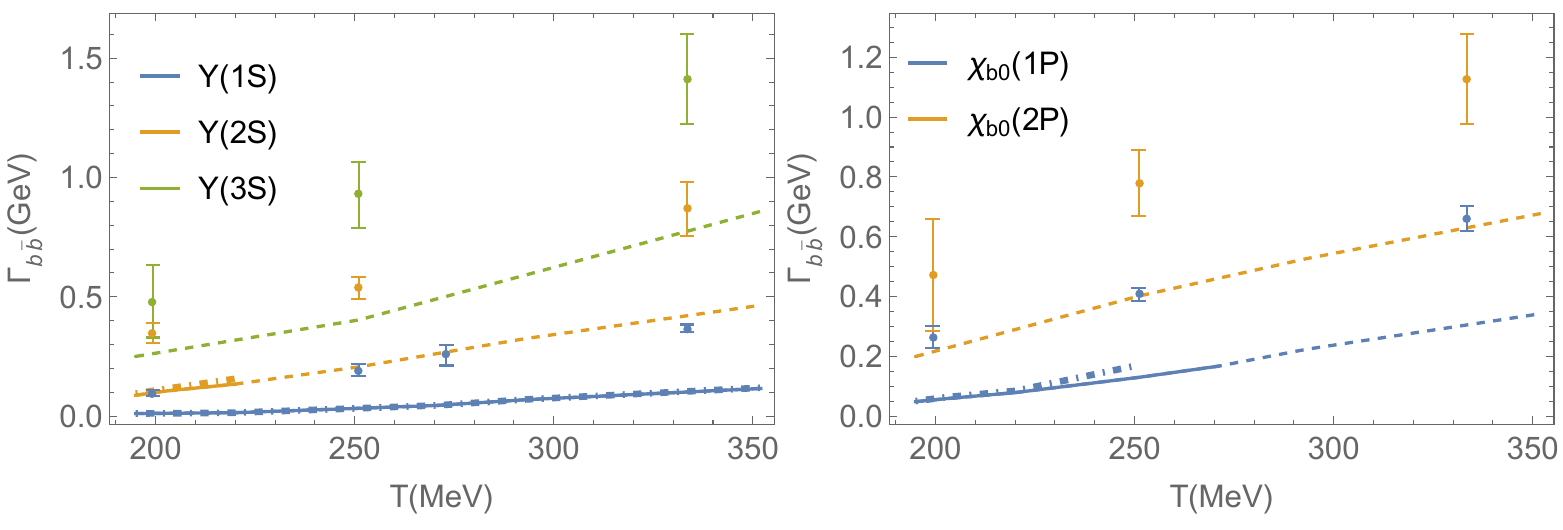}
\caption{Comparison of the widths from the spectral functions with extended (solid and dashed lines) and point operators (dot-dashed lines), as well as those extracted from the correlators with extended operators in a lattice study using a Gaussian form of the spectral function~\cite{Larsen:2019zqv,Larsen:2019bwy} (symbols with error bars), as a function of temperature for different bottomonium states. The dashed lines indicate the thermal widths of bottomonium states above the melting temperature, determined by whether the condition $\operatorname{det} \mathcal{F}(E)=0$ is satisfied. Left panel: $S$-wave states, right panel: $P$-waves states.} 
\label{fig_width}
\end{figure*}
To better understand the physical meaning of different structures in the spectral function 
and to facilitate the comparison with the lattice results it is useful
to quantify the width of these structures by extracting the full width at half maximum (FWHM), defined as the difference between the two energy values at which the spectral function drops to half of its peak height.
For spectral functions corresponding to the point operators this definition cannot always be executed since the right side of the peak merges with the continuum for high temperatures.
The widths for different states obtained from the spectral functions of extended meson operators are shown in Fig.~\ref{fig_width} as a function of temperature either as solid lines or dashed lines, where the latter continue beyond the melting temperature (\ie, where the poles on the real axis cease to exist).
The widths obtained from the spectral functions with point meson operators are displayed as dashed-dotted lines up to a temperature where the FWHM can be defined. We find reasonably good agreement between the results from the spectral functions corresponding to extended operators and the ones from point operators. 
In Fig.~\ref{fig_width} we also display the width parameter extracted from lattice calculations using a Gaussian parametrization of the peak of the spectral function~\cite{Larsen:2019zqv,Larsen:2019bwy}.
The FWHM is related to the Gaussian width parameter $\Gamma_G$ as FWHM$=2 \Gamma_G \sqrt{2\ln2}\simeq 2.355\Gamma_G$, which we take into account when comparing the $T$-matrix and lattice results for the width in Fig.~\ref{fig_width}.
The widths of different bottomonia states extracted from the fits in Refs.~\cite{Larsen:2019zqv,Larsen:2019bwy} are larger than the ones obtained in the $T$-matrix calculations. The main reason for this difference is the use of a simple Gaussian form
of the spectral function in Refs.~\cite{Larsen:2019zqv,Larsen:2019bwy}. For a Gaussian form of the spectral function the $\tau$
dependence of the effective masses is entirely determined by $\Gamma_G$, because the energy region far away from the peak does not contribute to the correlator, while for realistic spectral shapes the behavior
of the spectral function far away from the peak position is important for the $\tau$-dependence of the effective
masses. This is further discussed in Appendix~\ref{sec_sfex2},
where we also show that a cut Lorentzian form of the spectral function with a width parameter $\Gamma$, \ie, a form where the spectral function is
given by a Lorentzian (Brei-Wigner) in the interval $[M_{b \bar b}-2 \Gamma,M_{b \bar b}+2 \Gamma]$ with $M_{b \bar b}$ representing the peak position and is zero otherwise, provides a better parametrization of the spectral function from the lattice studies.

We thus conclude that the results from spectral functions with extended operators have to be interpreted with care. While for tightly bound states the results
are in agreement for both the peak mass and widths with the point spectral functions, 
this is no longer the case for the excited states. The extended operator results still produce peaks which however are not a reliable indication of a bound state. 
On the other hand, the ``masses" and ``widths" of these ``pseudo-peaks" can still provide information on the underlying potential and inelastic scattering rates of the heavy quarks at the distance probed by the optimized wave function.
Furthermore, using a Gaussian form of the spectral function when fitting the lattice results on the correlation function
tends to overestimate the in-medium bottomonium width appreciably.

%


\section{Heavy-Quark Transport Coefficients}
\label{sec_transport}
In this section, we utilize our newly constrained potential to predict HQ transport coefficients in the QGP. The main ingredients to the charm-quark transport coefficients are the heavy-light scattering amplitudes and parton spectral functions. 
Similar to our previous study~\cite{Tang:2023tkm}, the  parton spectral functions show strongly broadened peaks and collective modes on the low-energy shoulder of the quasiparticle peaks at zero parton momentum, even at high temperatures (which is different from the HQ free energy-based results~\cite{Liu:2017qah} where the interaction strength is more strongly screened at high temperatures). 
Another feature is that the parton widths do not fall off with 3-momentum as much as the ones from earlier potentials which focused on fitting the in-medium  HQ free energies; this is mainly due to the vector admixture to the confining interaction which generates relativistic (magnetic) corrections at large momenta (recall that a mixing coefficient of $\chi=0.8$ has been used as introduced in Sec.~\ref{sec_vac}). 
For the heavy-light scattering amplitudes, it is crucial to account for the broad spectral functions in the evaluation of the HQ transport coefficient~\cite{Liu:2018syc}, which enables to access the sub-threshold interaction strength generated by dynamically formed $D$-meson bound states at relatively low temperatures. 
Again, when comparing to the HQ free energy-based results~\cite{Liu:2017qah}, we find a harder momentum dependence of the scattering amplitudes as well as larger scattering amplitudes at high temperatures. 
However, our results do not differ much from those in Ref.~\cite{Tang:2023tkm} where the in-medium potentials have been constrained by static WLCs and QGP EoS but not by the bottomonium correlators with extended operators. This indicates that the systematic application of lQCD constraints stabilizes in terms of the underlying potentials and resulting transport properties of the QGP.



The HQ friction coefficient (relaxation rate) with off-shell spectral functions can be evaluated based on the Kadanoff-Baym equations as~\cite{Liu:2018syc}
\begin{equation}
\begin{aligned}
A(p)=& \sum_{i} \frac{1}{2 \varepsilon_{c}(p)} \int \frac{d \omega^{\prime} d^{3} \mathbf{p}^{\prime}}{(2 \pi)^{3} 2 \varepsilon_{c}\left(p^{\prime}\right)} \frac{d \nu d^{3} \mathbf{q}}{(2 \pi)^{3} 2 \varepsilon_{i}(q)} \frac{d \nu^{\prime} d^{3} \mathbf{q}^{\prime}}{(2 \pi)^{3} 2 \varepsilon_{i}\left(q^{\prime}\right)} \\
& \times \delta^{(4)} \frac{(2 \pi)^{4}}{d_{c}} \sum_{a,l, s}|M|^{2} \rho_{c}\left(\omega^{\prime}, p^{\prime}\right) \rho_{i}(\nu, q) \rho_{i}\left(\nu^{\prime}, q^{\prime}\right) \\
& \times \left[1-n_{c}\left(\omega^{\prime}\right)\right] n_{i}(\nu)\left[1 \pm n_{i}\left(\nu^{\prime}\right)\right] (1-\frac{\mathbf{p}\cdot\mathbf{p'}}{\mathbf{p}^2}) \ .
\end{aligned}
\label{eq_A(p)}
\end{equation}
Here,  $\delta^{(4)}$ denotes the energy-momentum conserving $\delta$-function and $d_c=6$ the spin-color degeneracy of charm quarks. The summation, $\sum_{i}$, is over all light anti-/quarks and gluons in the heat bath. The friction coefficient is evaluated for a definite energy and moment of the incoming charm quark, whereas all other partons are represented by off-shell spectral functions. The heavy-light scattering amplitudes, $|M|^2$, are related to the $T$-matrices and incorporate the summation over all possible two-body color and partial-wave channels as detailed in Ref.~\cite{Liu:2018syc}.

\begin{figure}[htbp]
\centering
\includegraphics[width=0.45\textwidth]{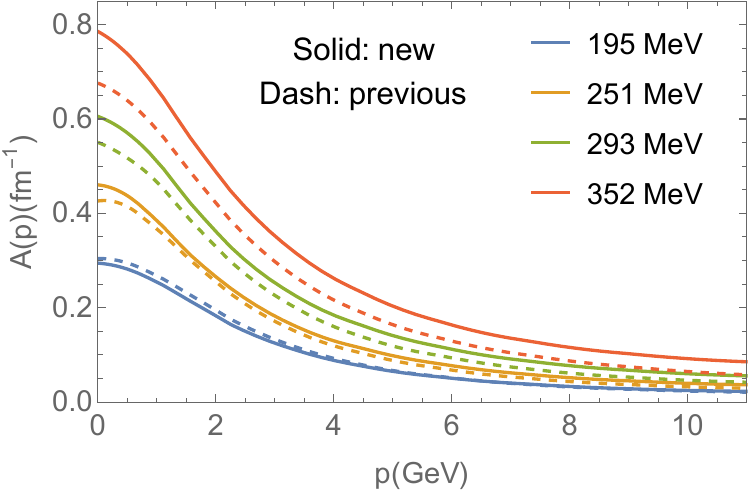}
\caption{The charm-quark friction coefficients in this (solid lines) and previous (dashed lines) studies as a function of 3-momentum at different temperatures.} 
\label{fig_Ap}
\end{figure}
Compared to our previous study~\cite{Tang:2023tkm}, where the in-medium potentials were constrained by the static WLCs and the QGP EoS, the charm-quark friction coefficient shown in Fig.~\ref{fig_Ap} is slightly larger at $T$=251, 293 and 334\,MeV, 
while at $T$=199\,MeV the results are very close to each other in both scenarios.
At low momenta, there is some compensation between the smaller charm-quark masses in the present work (which enhances $A(p)$ as the relaxation rate is proportional to the temperature-over-mass ratio), the stronger potential but also a stronger screening of the confining force ($m_s$=240\,MeV compared to 200\,MeV before).
At high momenta, where the short-range color-Coulomb interaction is dominant (further augmented by relativistic corrections), its weaker screening, especially at higher temperatures, implies that the $A(p)$ in this work is significantly larger than in the previous one.


The spatial diffusion coefficient, $D_s=T/(M_c A(p=0))$, scaled by the inverse thermal wavelength, $2\pi T$, is displayed in Fig.~\ref{fig_Ds} as a function of temperature.
Our results are in fair agreement with recent lQCD data~\cite{Altenkort:2023oms,Altenkort:2023eav} for charm and static quarks. The result for static quark is about a factor of 2-3 larger than results from the AdS/CFT correspondence  which are believed to provide a quantum lower bound for the diffusion coefficient~\cite{Casalderrey-Solana:2006fio}. 
Again, the present results are close to the WLC-based ones from Ref.~\cite{Tang:2023tkm}, although the $2\pi T D_s$ here is slightly higher at the lowest temperature due to a weaker (more screened) confining interaction.


\begin{figure}[htbp]
\centering
\includegraphics[width=0.45\textwidth]{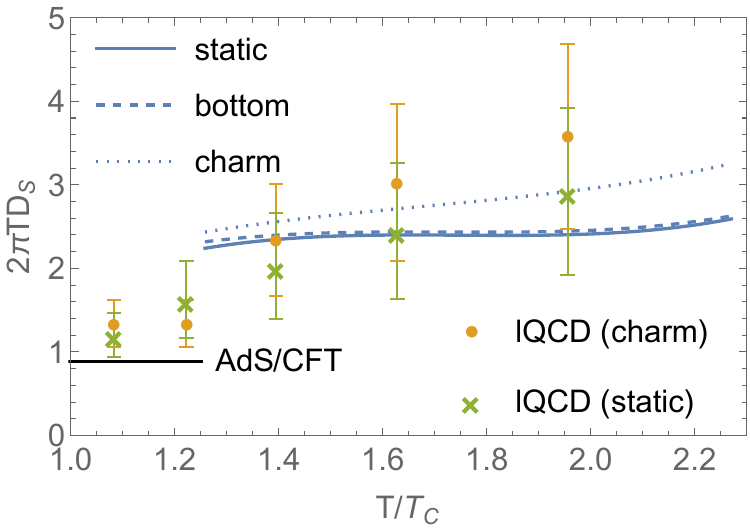}
\caption{The spatial diffusion coefficient for static (solid blue line), bottom (dashed blue line) and charm (dotted blue line) quarks as a function of temperature (scaled by $\Tpc$=155\,MeV) compared to the 2+1-flavor lQCD data~\cite{Altenkort:2023oms,Altenkort:2023eav} (yellow dots and green crosses for charm and static quarks, respectively) and an AdS/CFT estimate~\cite{Casalderrey-Solana:2006fio} (black line). The bare-quark masses for charm and bottom are 1.277 and 4.676~GeV, respectively, and we use 10\,GeV to approximate the bare static-quark mass.} 
\label{fig_Ds}
\end{figure}

\section{Conclusions}
\label{sec_concl}
We have employed the thermodynamic $T$-matrix approach to study bottomonium correlators with extended operators as recently computed in lattice QCD. Starting from a slightly amended vacuum potential (with larger coupling constant and string tension) which improves the bottomonium spectroscopy in vacuum, medium modifications have been constrained by the QGP equation of state, static Wilson line correlators and bottomonium correlators with extended operators from lQCD using selfconsistent $T$-matrix solutions. A fair agreement with lQCD data could be achieved for the 
effective masses of $S$- and $P$-wave bottomonia from the correlators with extended operators.
The spectral functions in the $T$-matrix approach with extended operators show peaks for all bottomonium states in the entire temperature region considered, even though some of these peaks
are very broad and asymmetric. 
The $S$- and $P$-wave spectral functions of point meson operators from the $T$-matrix approach feature a disappearance of the excited bottomonium states at certain temperatures. This suggests that not all peak structures in the spectral functions of extended operators can be interpreted as bound states, and the effective width obtained in lattice QCD calculations in Refs.~\cite{Larsen:2019bwy,Larsen:2019zqv} cannot be always interpreted as an in-medium width (or reaction rate).
From the poles of the $T$-matrix on the real energy axis utilizing Jost functions we assessed ``melting" temperatures of $3S$, $2P$, $2S$ and $1P$ bottomonium states to be $163,~174,~220$ and $293$ MeV, respectively. For $1S$ bottomonium state this temperature is larger than 334 MeV.
Well below the melting temperatures, the in-medium masses, determined as the peak position of the spectral functions of extended or point operators or as the real-axis pole of the $T$-matrix, agree with each other reasonably well. Furthermore, the in-medium widths obtained from the spectral functions of point and extended operators
also agree below the melting temperatures. This implies that while the spectral functions of extended meson operators cannot necessarily be used to study the melting of different bottomonium
states, they are still suitable to study in-medium bottomonium properties for smaller temperatures. We also note that the effective width obtained in the $T$-matrix
approach with extended meson operators  is smaller than the one obtained in lattice QCD calculations~\cite{Larsen:2019bwy,Larsen:2019zqv}. This is in large part due to the use of a simple Gaussian form of the spectral function, which does not capture the low-energy tails found in the 
microscopic calculations of the quantum many-body approach. With a more realistic form such as a Lorentzian form, the spectral function in the lattice analysis would yield smaller in-medium bottomonium width, although energy cutoffs are required to mimic the energy dependent selfenergies as 
generated by microscopic calculations.

The inferred in-medium potentials are quite similar to those in our previous study where they have been constrained by the QGP EoS and static WLCs only, but the interference effects turn out to be more significant when also including the constraints from correlators with extended operators. The main features for the spectral and transport properties from our previous and current studies remain intact: significant low-energy collective modes for parton spectral functions at high temperatures, resonant structures in the heavy-light scattering amplitudes, and a weak temperature dependence of the predicted spatial diffusion coefficient in fair agreement with recent 2+1-flavor lQCD results. The thermal relaxation rate (friction coefficient) in the present work is slightly larger 
at higher temperatures but closely agrees with our previous study at lower ones.

Ongoing work is directed at implementing these transport coefficients into phenomenological applications in heavy-ion collisions, to evaluate the impact on, and theoretical uncertainty in, describing the open HF observables. This is, in fact, part of a larger effort to utilize the nonperturbative $T$-matrices also in the quarkonium sector, \eg, to calculate their inelastic reaction rates and the pertinent (quantum) kinetics of heavy quarkonia in an evolving QCD medium.  

\acknowledgments
This work has been supported by the U.S. National Science Foundation under grant no. PHY-2209335, by the U.S. Department of Energy, Office of Science, Office of Nuclear Physics through contract No. DE-SC0012704 and the Topical Collaboration in Nuclear Theory on \textit{Heavy-Flavor Theory (HEFTY) for QCD Matter} under award no.~DE-SC0023547. We thank Hai-Tao Shu and Rasmus Larsen for helpful discussions.

 \appendix

\section{Spectral Representation for Quarkonium Correlators with Extended Operators}
\label{sec_sfex}
In this section, we derive the spectral representation for the quarkonium correlators with extended operators following Refs.~\cite{Karsch:2000gi,Aarts:2005hg,Alberico:2004we}.
The summation in Eq.~(\ref{Oext}) becomes an integration in the continuum limit:
\begin{equation}
O_i(\mathbf{x}, t)=\frac{1}{V}\int d^3\mathbf{r} \Psi_i(\mathbf{r}) \bar{q}(\mathbf{x}+\mathbf{r}, t) \Gamma q(\mathbf{x}, t) \ ,
\end{equation}
where $V$ is the lattice volume. The correlator introduced in Sec.~\ref{ssec_ext-op} is connected to its frequency-momentum-space components by a Fourier transform,
\begin{equation}
G_{i j}( \mathbf{x},\tau)=\frac{1}{\beta} \sum_{n=-\infty}^{\infty} \int \frac{d^3 \mathbf{P}}{(2 \pi)^3} e^{-i\left(\omega_n \tau-\mathbf{P} \cdot \mathbf{x}\right)} G_{i j}\left(\mathbf{P},i \omega_{n}\right) 
\ ,
\label{eq_Gtx}
\end{equation}
where $\mathbf{P}$ is the total momentum of bottomonium which is usually set to $\mathbf{P}=0$. Defining the bottomonium spectral functions as 
\begin{equation}
\rho_{Q\bar{Q}}^{ij}( \mathbf{P},E)=-\frac{1}{\pi} \operatorname{Im} G_{ij}(\mathbf{P},E+i \eta) \ ,
\label{eq_rhoEP}
\end{equation}
the correlators can be expressed in a mixed representation through the spectral functions as
\begin{equation}
G_{ij}(\mathbf{P},\tau)=\int_{-\infty}^{\infty} d E \rho_{Q\bar{Q}}^{ij}(\mathbf{P},E) e^{-E \tau} \ .
\label{eq_GtP}
\end{equation}
To obtain the correlators, $G_{ij}(\mathbf{P},\tau)$, required for evaluating the effective mass in Eq.~(\ref{eq_Meff}), one has to first find $G_{ij}\left(\mathbf{P},i \omega_n\right)$ through the inverse transform of Eq.~(\ref{eq_Gtx}):
\begin{equation}
G_{ij}\left( \mathbf{P},i \omega_n\right)=\int d^3 \mathbf{x} \int_0^\beta d \tau e^{i\left(\omega_n \tau-\mathbf{P} \cdot\mathbf{x}\right)} G_{ij}( \mathbf{x},\tau) \ 
\end{equation}
and then plug it into Eq.~(\ref{eq_rhoEP}) to yield the spectral functions. Finally, the correlators, $G_{ij}( \mathbf{P},\tau)$, can be computed via Eq.~(\ref{eq_GtP}).

The correlation function, $G_{ij}( \mathbf{x},\tau)$, consists of two components, \ie, the free part, ${G}_{ij}^0(\mathbf{x},\tau)$, and the interacting part, $\Delta {G}_{ij}(\mathbf{x},\tau)$, which can be evaluated directly from the definition introduced in 
Sec.~\ref{ssec_ext-op}. For the free contribution, one has
\begin{equation}
\begin{aligned}
& G_{ij}^0( \mathbf{x},\tau)=\left\langle O_i( \mathbf{x},\tau) O_j^{\dagger}( \mathbf{0},0)\right\rangle \\
& =\left\langle\sum_{\mathbf{r}, \mathbf{r}^{\prime}} \Psi_i(\mathbf{r}) \bar{q}( \mathbf{x}+\mathbf{r},\tau) \Gamma q(\mathbf{x},\tau) \Psi_j\left(\mathbf{r}^{\prime}\right) \bar{q}(\mathbf{0},0) \Gamma q(\mathbf{r},0)\right\rangle \\
& =\frac{N_f N_c }{V^2}\int d^3 \mathbf{r} d^3 \mathbf{r}^{\prime} \Psi_i(\mathbf{r}) \Psi_j(\mathbf{r}) \\
& \times \operatorname{Tr}\left[\Gamma S_F( \mathbf{x},\tau) \Gamma S_F\left( \mathbf{r}^{\prime}-\mathbf{r}-\mathbf{x},\beta-\tau\right)\right] \ ,
\end{aligned}
\end{equation}
where $S_F$ is the quark propagator~\cite{Alberico:2004we}. Then, the correlators in $(\mathbf{P},i \omega_n)$ space are given by
\begin{widetext}
\begin{equation}
\begin{aligned}
& {G}_{ij}^0\left(\mathbf{P}, i \omega_n\right)=\int d^3 \mathbf{x} \int_0^\beta d \tau e^{i\left(\omega_n \tau-\mathbf{P} \cdot\mathbf{x}\right)} {G}_{ij}^0(\mathbf{x}, \tau) \\
& =\frac{N_f N_c}{V^2} \int_0^\beta d \tau e^{i \omega_n \tau} \int \frac{d^3 \mathbf{k}_1}{(2 \pi)^3} \frac{d^3 \mathbf{k}_2}{(2 \pi)^3} \frac{d^3 \mathbf{k}_3}{(2 \pi)^3} \frac{d^3 \mathbf{k}_4}{(2 \pi)^3} {\Psi}_i\left(\mathbf{k}_1\right) {\Psi}_j\left(\mathbf{k}_2\right) \operatorname{Tr}\left[\Gamma S_F\left(\mathbf{k}_3, \tau\right) \Gamma S_F\left(\mathbf{k}_4, \beta-\tau\right)\right] \\
& \quad \times(2 \pi)^3 \delta\left(-\mathbf{P}+\mathbf{k}_3-\mathbf{k}_4\right)(2 \pi)^3 \delta\left(\mathbf{k}_1-\mathbf{k}_4\right)(2 \pi)^3 \delta\left(\mathbf{k}_2+\mathbf{k}_4\right) \\
& =\frac{N_f N_c}{V^2} \int_0^\beta d \tau e^{i \omega_n \tau} \int \frac{d^3 \mathbf{k}}{(2 \pi)^3}  {\Psi}_i(\mathbf{k})  {\Psi}_j(\mathbf{k}) \operatorname{Tr}\left[\Gamma S_F(\mathbf{P}+\mathbf{k}, \tau) \Gamma S_F(\mathbf{k}, \beta-\tau)\right] \\
& =\frac{N_f N_c}{V^2} \int \frac{d^3 \mathbf{k}}{(2 \pi)^3}  {\Psi}_i(\mathbf{k})  {\Psi}_j(\mathbf{k}) \int_0^\beta d \tau \frac{1}{\beta^2} \sum_{n_1, n_2=-\infty}^{+\infty} e^{i\left(\omega_n-\omega_{n_1}+\omega_{n_2}\right) \tau} \operatorname{Tr}\left[\Gamma S_F\left(\mathbf{P}+\mathbf{k}_{,} i \omega_{n_1}\right) \Gamma S_F\left(\mathbf{k}, i \omega_{n_2}\right)\right] \\
& =\frac{N_f N_c}{V^2} \int \frac{d^3 \mathbf{k}}{(2 \pi)^3}  {\Psi}_i(\mathbf{k})  {\Psi}_j(\mathbf{k}) \frac{1}{\beta} \sum_{n=-\infty}^{+\infty} \operatorname{Tr}\left[\Gamma S_F\left(\mathbf{P}+\mathbf{k}, i \omega_{n_1}\right) \Gamma S_F\left(\mathbf{k}, i \omega_{n_1}-i \omega_n\right)\right] \\
& =\frac{N_f N_c}{V^2} \int \frac{d^3 \mathbf{k}}{(2 \pi)^3}  {\Psi}_i(\mathbf{k})  {\Psi}_j(\mathbf{k}) \mathcal{R}_{Q\bar{Q}}^{sca}G_{Q \bar{Q}}^0(E, \mathbf{k}) \operatorname{Tr}\left[\Gamma \Lambda_{+}(\mathbf{k}) \Gamma \Lambda_{-}(-\mathbf{k})\right]\\
& =\frac{N_f N_c}{V^2}\int \frac{d^3k}{(2\pi )^3}\mathcal{R}_{Q\bar{Q}}^{sca} a_{free}(k) G_{Q\bar{Q}}^0(E,k) {\Psi}^2(k) \ ,
\end{aligned}
\label{eq_G0}
\end{equation}
\end{widetext}

where the Fourier transforms, $\Psi_i(\mathbf{r})= \int \frac{d^3 \mathbf{k}}{(2 \pi)^3} e^{i \mathbf{k} \cdot \mathbf{r}}  {\Psi}_i(\mathbf{k})$ and $S_F(\mathbf{x}, \tau)=\int \frac{d^3 \mathbf{k}}{(2 \pi)^3} e^{i \mathbf{k} \cdot \mathbf{x}} S_F(\mathbf{k}, \tau)$, have been applied in the second equality, and $S_F(\mathbf{k}, \tau)=\frac{1}{\beta} \sum_{n=-\infty}^{+\infty} e^{-i \omega_n \tau} S_F\left(\mathbf{k}, i \omega_n\right)$ in the fourth equality. In the fifth equality of Eq.~(\ref{eq_G0}), we have used the fact that the Matsubara frequencies for mesons and quarks are $\omega_n=2n\pi T$ and $\omega_n=2(n+1)\pi T$, respectively. A positive-energy projected propagator has been used in the sixth equality: $G_{Q \bar{Q}}^0$ is the two-body propagator introduced in Eq.~(\ref{G2}), and $\Lambda_{ \pm}(\mathbf{k})=\left[\varepsilon_Q(k) \gamma^0-(\mathbf{k} \cdot \boldsymbol{\gamma}) \pm M_Q\right] / 2 M_Q$ are the positive/negative energy projectors~\cite{Cabrera:2006wh,Riek:2010fk,Tang:2023lcn} for quark ($+$) and antiquark ($-$), respectively.

The interacting part of correlators is given by:
\begin{widetext}
\begin{equation}
\begin{aligned}
& \Delta  {G}_{ij}(\mathbf{x},\tau)=\left\langle O_i(\mathbf{x}, \tau) J\left(\mathbf{x}^{\prime}, \tau^{\prime}\right) \mathcal{T} J^{\dagger}\left(\mathbf{x}^{\prime}, \tau^{\prime}\right) O_j^{\dagger}(\mathbf{x}, 0)\right\rangle \\
& =\frac{1}{V^2} \int d^3 \mathbf{x}^{\prime} d^3 \mathbf{r} d^3 \mathbf{r}^{\prime} \int d \tau^{\prime} \left\langle\Psi_i(\mathbf{r}) \bar{q}(\mathbf{x}+\mathbf{r}, \tau) q(\mathbf{x}, \tau) \bar{q}\left(\mathbf{x}^{\prime}, \tau^{\prime}\right) q\left(\mathbf{x}^{\prime}, \tau^{\prime}\right) \mathcal{T} \bar{q}\left(\mathbf{x}^{\prime}, \tau^{\prime}\right) q\left(\mathbf{x}^{\prime}, \tau^{\prime}\right) \Psi_j\left(\mathbf{r}^{\prime}\right) \bar{q}\left(\mathbf{x}^{\prime}, 0\right) q\left(\mathbf{r}^{\prime}, 0\right)\right\rangle \\
& =\frac{N_f N_c}{V^2} \int d^3 \mathbf{x}^{\prime} d^3 \mathbf{r} d^3 \mathbf{r}^{\prime} \int d \tau^{\prime} \Psi_i(\mathbf{r}) \Psi_j\left(\mathbf{r}^{\prime}\right) \mathcal{T} \\
& \times \operatorname{Tr}\left[S_F\left(\mathbf{x}-\mathbf{x}^{\prime}, \tau-\tau^{\prime}\right) S_F\left(\mathbf{x}^{\prime}-\mathbf{x}-\mathbf{r}, \beta+\tau^{\prime}-\tau\right) S_F\left(\mathbf{x}^{\prime}-\mathbf{x}-\mathbf{r}, \tau^{\prime}\right) S_F\left(\mathbf{r}^{\prime}-\mathbf{x}^{\prime}, \beta-\tau^{\prime}\right)\right],
\end{aligned}
\end{equation}
\end{widetext}

where $ J(\mathbf{x}, \tau)=\bar{q}(\mathbf{x}, \tau)q(\mathbf{x}, \tau)$ is the point operator, and $\mathcal{T}$ denotes the 2-body scattering amplitude. Applying similar techniques as used in Eq.~(\ref{eq_G0}), the correlators in $(\mathbf{P},i \omega_n)$ space are given by
\begin{widetext}
\begin{equation}
\begin{aligned}
& \Delta {G}_{ij}\left(\mathbf{P}, i \omega_n\right)=\int d^3 \mathbf{x} \int_0^\beta d \tau e^{i\left(\omega_n \tau-\mathbf{P} \cdot\mathbf{x}\right)} \Delta {G}_{ij}(\mathbf{x}, \tau) \\
& =\frac{N_f N_c}{V^2} \int \frac{d^3 \mathbf{k}}{(2 \pi)^3}\mathcal{R}_{Q\bar{Q}}^{sca} G_{Q \bar{Q}}^0(E, \mathbf{k}) {\Psi}_i(\mathbf{k})  \int \frac{d^3 \mathbf{k}'}{(2 \pi)^3} G_{Q \bar{Q}}^0(E, \mathbf{k'}) {\Psi}_j(\mathbf{k}')\mathcal{T}(E,\mathbf{k},\mathbf{k}')\operatorname{Tr}\left[\Lambda_{+}(\mathbf{k})\Gamma\Lambda_{-}(-\mathbf{k})\Lambda_{-}(-\mathbf{k'})  \Gamma\Lambda_{+}(\mathbf{k'})\right]\\
& =\frac{N_f N_c}{V^2\pi ^3}\int dk k^2 \mathcal{R}_{Q\bar{Q}}^{sca} G_{Q\bar{Q}}^0(E,k)\int dk' k'^2 G_{Q\bar{Q}}^0(E,k') [a_0(k,k')T_{Q\bar{Q}}^{0}+a_1(k,k')T_{Q\bar{Q}}^{1}] {\Psi}(k) {\Psi}(k').
\end{aligned}
\label{eq_deltaG}
\end{equation}
\end{widetext}

Combining  Eq.~(\ref{eq_G0}) and Eq.~(\ref{eq_deltaG}) gives 
Eq.~(\ref{eq_Cmed}) (recall that the total momentum is set to $\mathbf{P}=0$). The overall factor $1/V^2$ is canceled out in the ratio used to define the effective mass, see Eq.~(\ref{eq_Meff}). The derivation presented above, which uses non-optimized operators, can be readily extended to the one with optimized operators, given that the latter are simply linear combinations of the former as discussed in 
Sec.~\ref{ssec_ext-op}.

\section{Spectral Functions for Point- and Extended-Meson Operators}
\label{sec_sfex2}
In this appendix we discuss some details of the determination of the thermal widths from the spectral functions for point and extended meson operators and the comparison with the lattice QCD results. 
In Fig.~\ref{fig_SFbb_ext_vs_point} we show the bottomonium spectral function of $\Upsilon$ states calculated with point and 
extended meson operators using logarithmic scale on the $y$-axis. 
We see that for energies, $E$, well below the peak position the spectral function decays exponentially for both types
of meson operators. This exponential decrease is related to the exponential decrease of the imaginary part of the in-medium $b$-quark selfenergy at small energies, as shown in Fig.~\ref{fig_selfenergy_b} for $T$=199\,MeV as an example.
In the case of point operators the spectral function does not decrease
with increasing $E$ for energies well above the peak. For extended meson
operators the spectral function decreases with increasing $E$ for sufficiently large $E$ values, but we see a prominent shoulder structure above the peak, which leads to significant increase in the effective mass at small $\tau$. 
\begin{figure*}[tbp]
\begin{minipage}[b]{1.0\linewidth}
\centering
\includegraphics[width=0.99\textwidth]{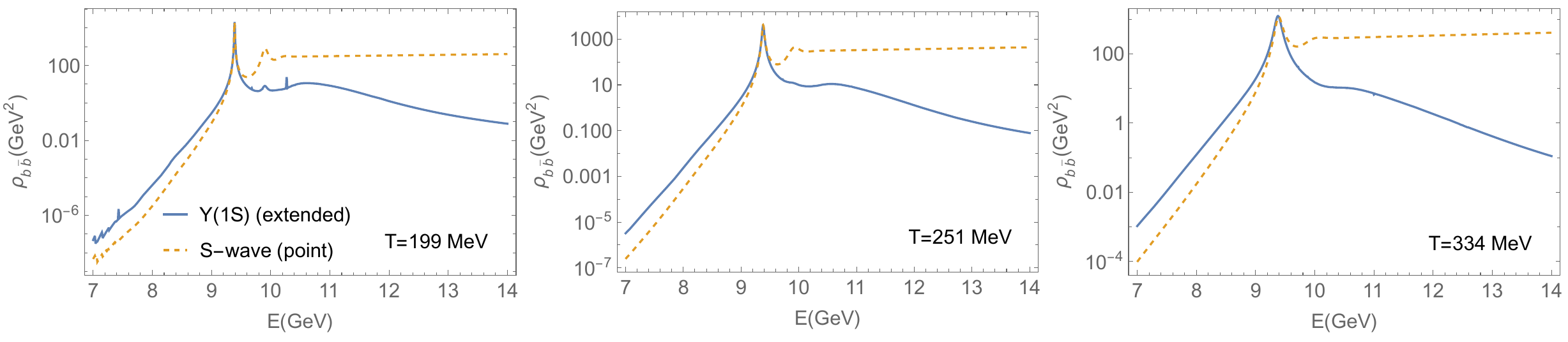}
\end{minipage}
\caption{The $\Upsilon(1S)$ bottomonium spectral functions with extended operators (solid lines) and the $S$-wave bottomonium spectral functions with point operators (dashed lines) as a function of energy. The spectral functions with point operators have been scaled up by suitable factors to be comparable to those with extended operators.} 
\label{fig_SFbb_ext_vs_point}
\end{figure*}
\begin{figure*}[tbp]
\begin{minipage}[b]{1.0\linewidth}
\centering
\includegraphics[width=0.5\textwidth]{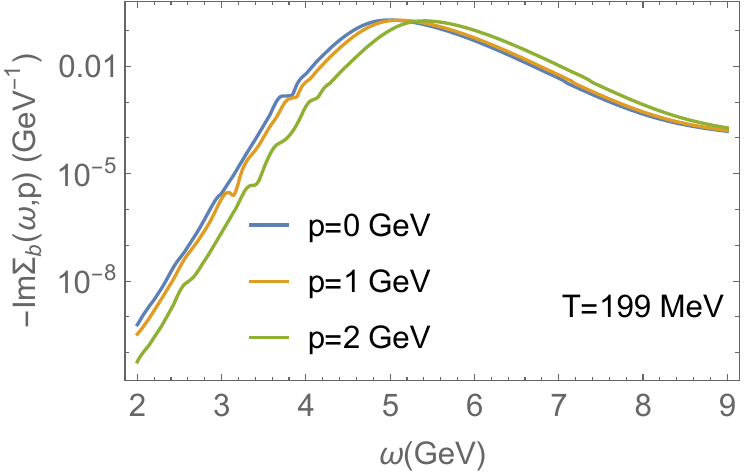}
\end{minipage}
\caption{The imaginary part of the quark selfenergy as function of 
$\omega$ at $T=199$ MeV for three different spatial momenta.} 
\label{fig_selfenergy_b}
\end{figure*}
In Fig. \ref{fig_SFbb_Tm_vs_GS} we show the spectral function corresponding to the extended-meson operators for $1S$, $2S$, $3S$, $1P$ and $2P$ states. We see that for all states the spectral function decays exponentially far away from the peak at about the same rate, although there is a shoulder structure visible for the $1S$ and $2S$ spectral functions.
It is instructive to compare the obtained spectral function with a simple Gaussian parametrization, 
\begin{align}
A_{\mathrm{G}}(T) \exp \left(-\frac{\left[\omega-M_{b \bar b}(T)\right]^{2}}{2 \Gamma_{\mathrm{G}}^{2}(T)}\right) \ ,
\label{eq:Gaussian}	
\end{align}
with a Gaussian parameter determined from the FWHM (see main text), and $M_{b \bar b}(T)$ being the in-medium bottomonium mass. 
This parametrization is also shown in  Fig.~\ref{fig_SFbb_Tm_vs_GS} as dashed lines. While in the vicinity of the peak the Gaussian 
parametrization describes the spectral function fairly well, it is 
very different away from the peak, \ie, it is much smaller than the spectral
function from the $T$-matrix calculations.
In Fig. \ref{fig_SFbb_Tm_vs_GS_lattice} we compare our spectral functions
corresponding to extended meson operators for different states with the Gaussian parametrization used in Ref.~\cite{Larsen:2019zqv}. We see that the Gaussian parametrization used in the lattice study gives much broader peaks. This is in order to describe the $\tau$ dependence
of the lattice bottomonium correlators and compensate for the rapid fall-off of the Gaussian a larger width is needed. 

Next we compare our spectral function with the Breit-Wigner (Lorentzian) parametrization, 
\begin{align}
A(T) \frac{(\Gamma/2)}{(\omega-M_{b\bar b}(T))^2+\frac{\Gamma^2}{4}} \ ,
\end{align}
with $\Gamma$ being the FWHM and $M_{b \bar b}(T)$ the in-medium mass of the bottomonium state.
\begin{figure*}[tbp]
\begin{minipage}[b]{1.0\linewidth}
\centering
\includegraphics[width=0.99\textwidth]{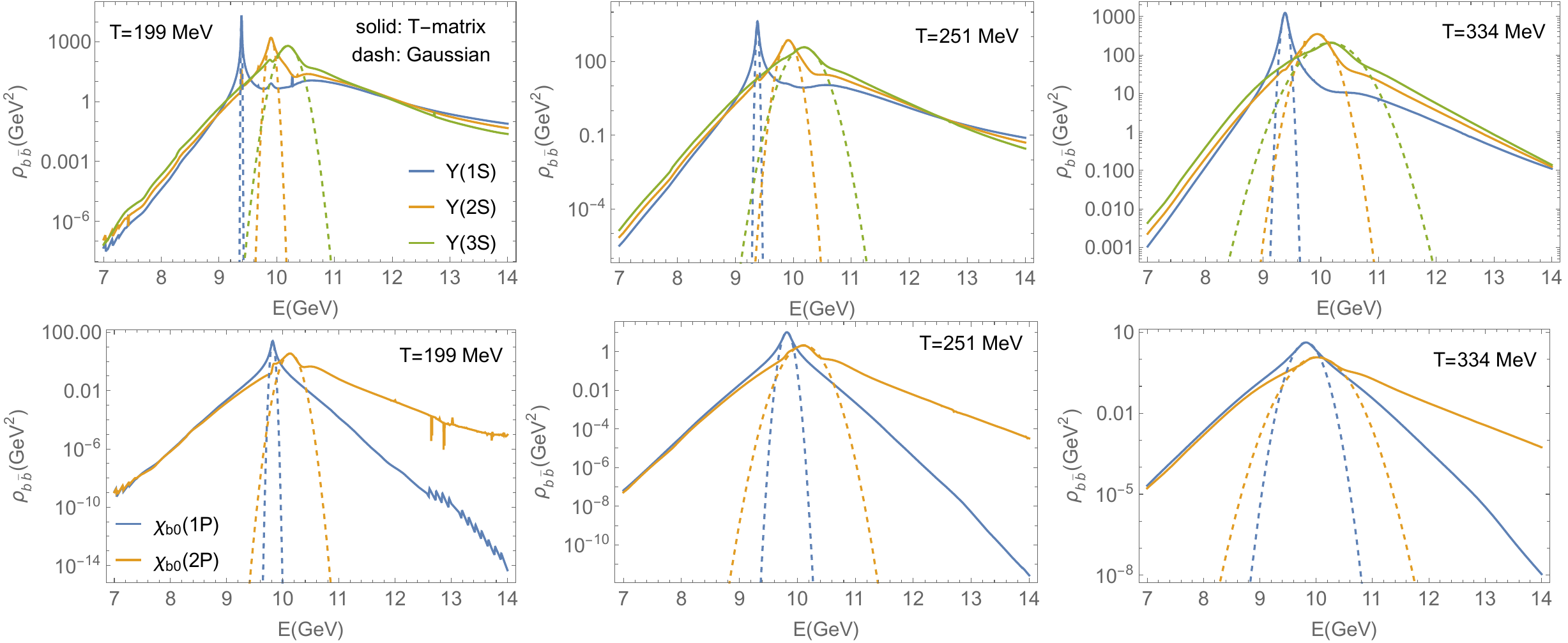}
\end{minipage}
\caption{The $\Upsilon$ (upper panels) and $\chi_{b0}$ (lower panels) bottomonium spectral functions from the $T$-matrix (solid lines) as a function of total energy, compared to spectral functions using a Gaussian parametrization (dashed lines) at different temperatures. The Gaussian spectral functions have been scaled up by a suitable factor to be comparable to the $T$-matrix results.} 
\label{fig_SFbb_Tm_vs_GS}
\end{figure*}
\begin{figure*}[tbp]
\begin{minipage}[b]{1.0\linewidth}
\centering
\includegraphics[width=0.99\textwidth]{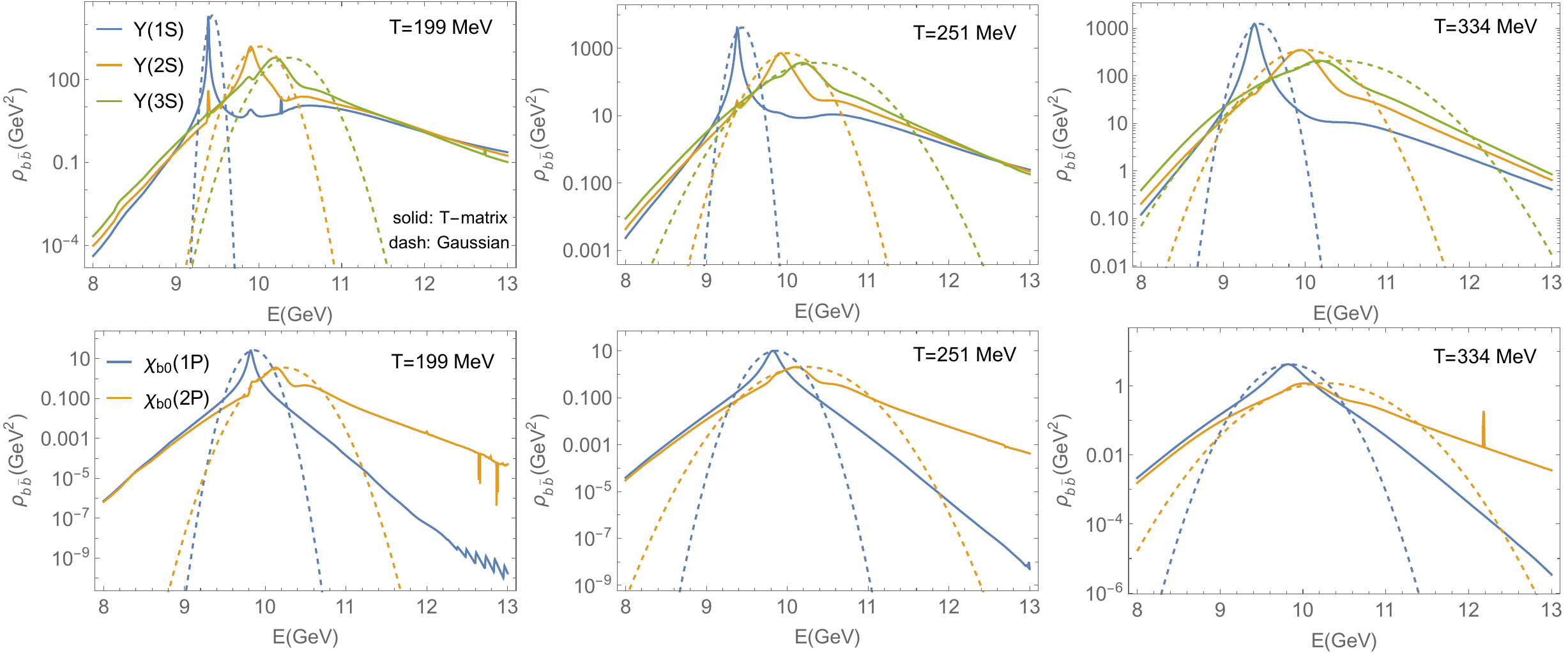}
\end{minipage}
\caption{Spectral functions for the $\Upsilon$ (upper panels) and $\chi_{b0}$ (lower panels) bottomonium states. Solid curves represent the $T$-matrix results as a function of total energy, while dashed curves show Gaussian parameterizations using mass and width values from lattice studies at varying temperatures. The Gaussian spectral functions are scaled for comparison with the $T$-matrix results.} 
\label{fig_SFbb_Tm_vs_GS_lattice}
\end{figure*}
In Figs.~\ref{fig_SFbb_Tm_vs_BW} and \ref{fig_SFbb_Tm_vs_BW2} we compare this parametrization with the spectral function obtained from the 
$T$-matrix calculations. We see that in the interval 
$[M_{b \bar{b}}-2\Gamma,M_{b \bar{b}}+2\Gamma]$ the Breit-Wigner form provides a good description of the full spectral function corresponding to 
the extended meson operators of $1S$, $2S$ and $1P$ states, and a fair description of the full spectral function corresponding to $3S$ and 
$2P$ states. The Laplace transform of the Lorentzian form does not
exist because the corresponding integrand does not vanish exponentially for small $E$. For this reason a Lorentzian form cannot be used to model 
the lattice QCD results, and a Gaussian form was used. However, if we introduce a cutoff into the Lorentzian spectral function, \ie, 
a Lorentzian in the interval $[M_{b \bar{b}}-2\Gamma,M_{b \bar{b}}+2\Gamma]$ and zero otherwise, then the Laplace transform can be performed. In Fig.~\ref{fig_Meff_Tm_vs_BW} we show the effective masses obtained from such a form of the spectral function and compare them
to the effective masses obtained from the $T$-matrix spectral function and the effective masses from lattice NRQCD calculations~\cite{Larsen:2019zqv}. We see that at small $\tau$ the effective masses obtained from the cut Lorentzian form are very similar to the effective mass obtained
with the $T$-matrix spectral function. There are differences in the effective masses at larger $\tau$. However, even the $T$-matrix spectral function cannot fully describe the lattice QCD results on the effective masses at large $\tau$. 
This is due to an additional contribution to the bottomonium spectral function at very low energy specific to NRQCD~\cite{Bala:2021fkm}, 
which is unrelated to the bound-state peak. Such a contribution is not present in the $T$-matrix approach or in lQCD calculations with relativistic heavy quarks. Thus, a cut Lorentzian form provides a reasonable parametrization of the spectral function that can be used in
lattice QCD analysis. Using such a parametrization for fitting the effective masses would result in significantly smaller width parameters, $\Gamma$, and the difference between the lattice and $T$-matrix results in 
Fig.~\ref{fig_width} is understood as being due to using an unrealistic form
of the spectral function in Ref.~\cite{Larsen:2019zqv}.

\begin{figure*}[tbp]
\begin{minipage}[b]{1.0\linewidth}
\centering
\includegraphics[width=0.99\textwidth]{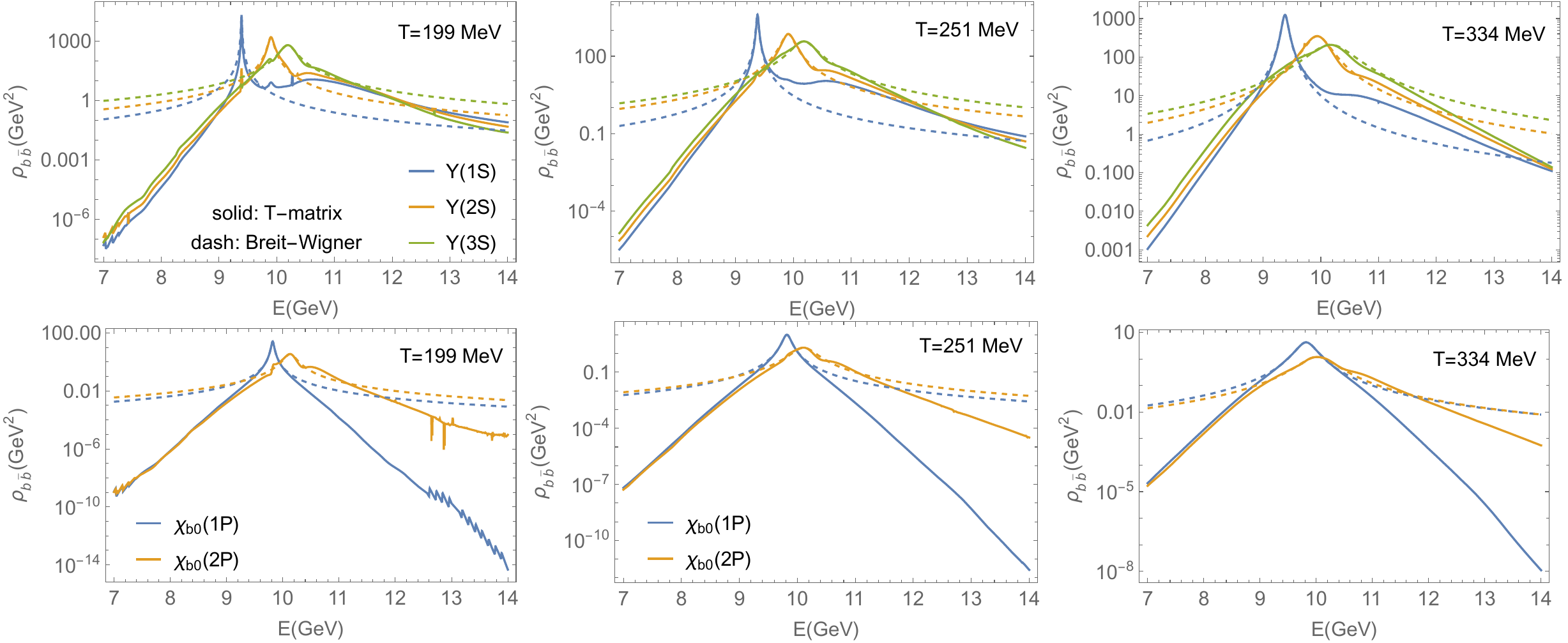}
\end{minipage}
\caption{The $\Upsilon$ (upper panels) and $\chi_{b0}$ (lower panels) bottomonium spectral functions from the $T$-matrix results (solid) as a function of total energy, compared to the spectral functions using a Breit-Wigner form parametrization (dashed) at different temperatures. The Breit-Wigner form spectral functions have been scaled up by certain factors to be comparable to those from $T$-matrix results.} 
\label{fig_SFbb_Tm_vs_BW}
\end{figure*}
\begin{figure*}[tbp]
\begin{minipage}[b]{1.0\linewidth}
\centering
\includegraphics[width=0.99\textwidth]{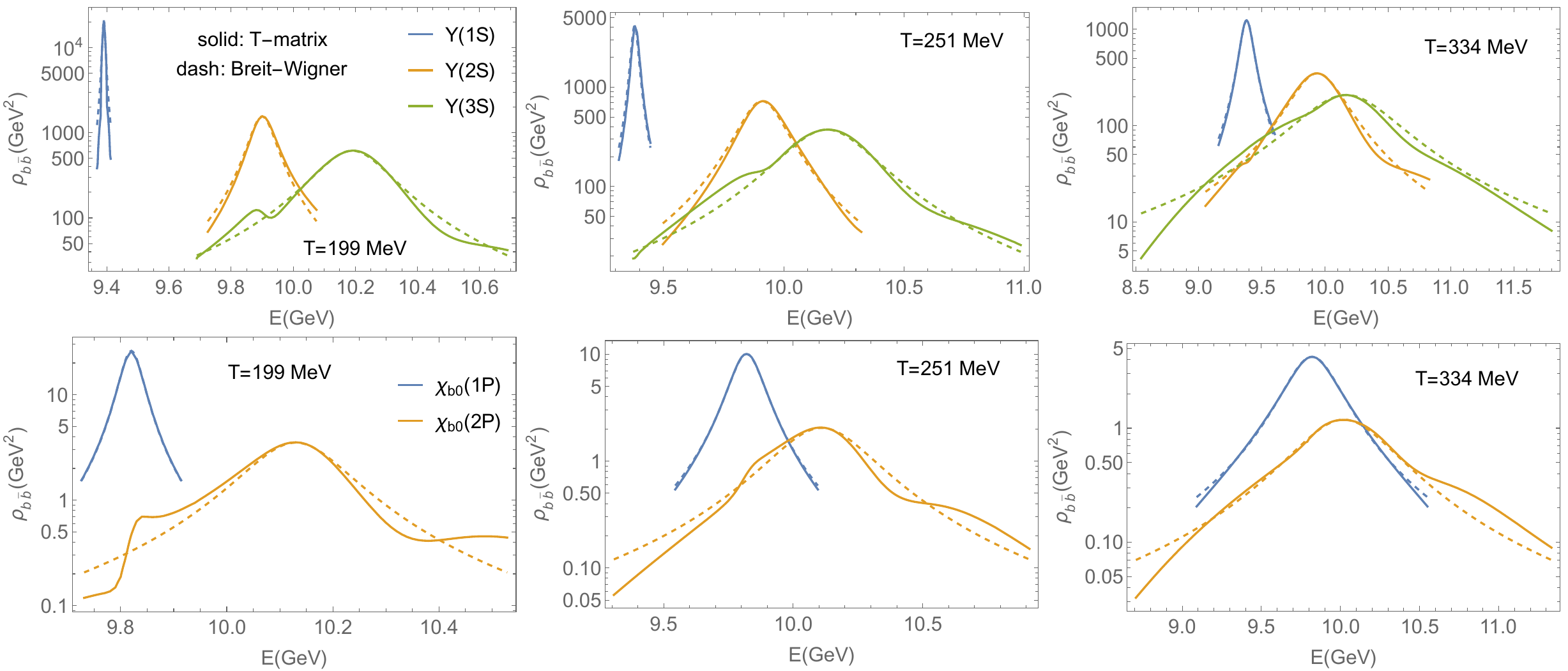}
\end{minipage}
\caption{Same as Fig.~\ref{fig_SFbb_Tm_vs_BW}, but restricted to the energy interval $[M_{b \bar{b}}-2\Gamma,M_{b \bar{b}}+2\Gamma]$, where $\Gamma$ represents the width as shown in Fig.\ref{fig_width}.} 
\label{fig_SFbb_Tm_vs_BW2}
\end{figure*}

 \begin{figure*}[tbp]
\centering
\includegraphics[width=0.97\textwidth]{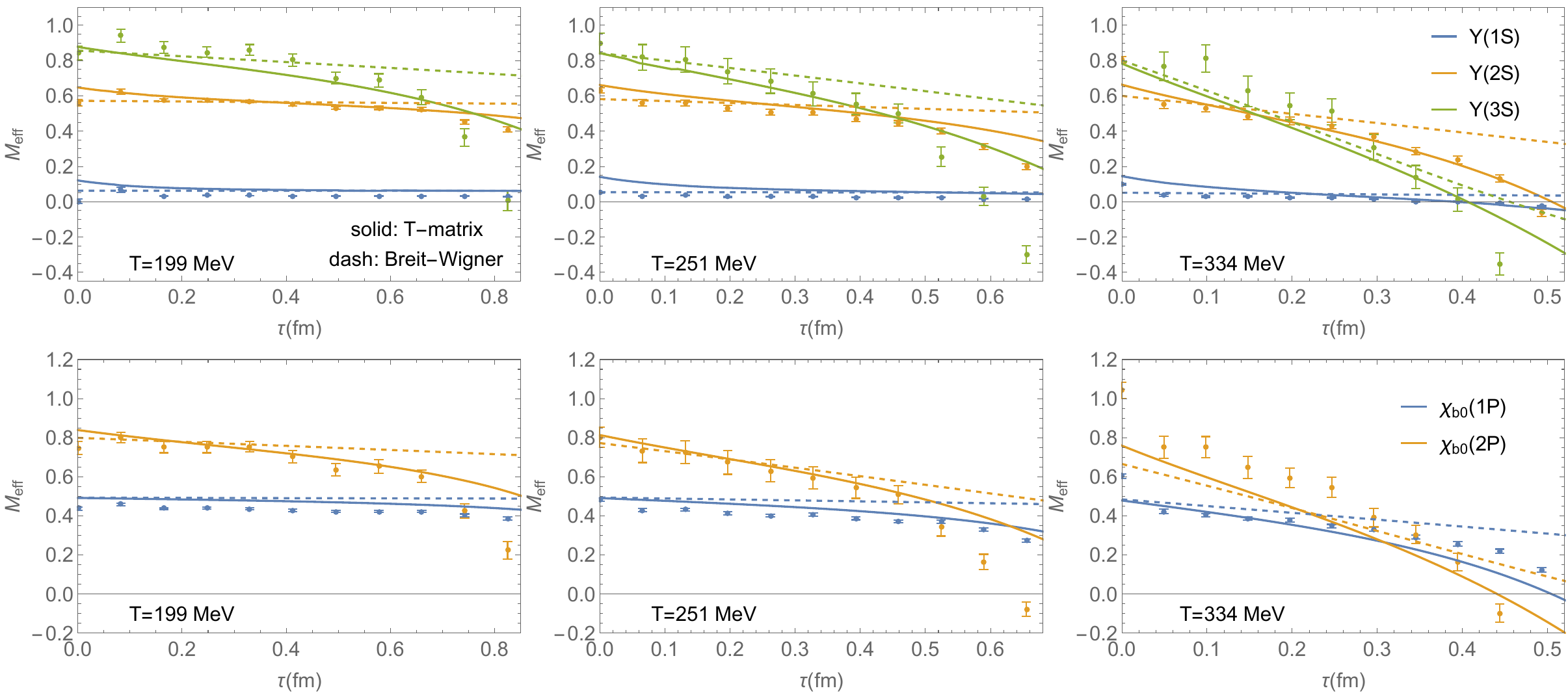}
\caption{The effective masses of bottomonium correlators with extended operators from the $T$-matrix (lines) and from the spectral functions using a Breit-Wigner form parameterization (dashed) as a function of imaginary time at different temperatures compared to the corresponding 2+1-flavor lQCD data~\cite{Larsen:2019zqv} (dots). Upper panel: the effective masses for $\Upsilon(1S)$ (blue), $\Upsilon(2S)$ (orange) and $\Upsilon(3S)$ (green) states. Lower panel: the effective masses for $\chi_{b0}(1P)$ (blue) and $\chi_{b0}(2P)$ (orange) states.} 
\label{fig_Meff_Tm_vs_BW}
\end{figure*}

\section{Potential Comparisons to Existing Work}
\label{sec_sfex3}
In this appendix, we compare both the real and imaginary parts of the static potential from our work to the findings of other studies in the literature. Recall that the real part is the screened Cornell potential, given by Eq.~(\ref{eq_Vr}), with parameters shown in Fig.~\ref{fig_V}. The imaginary part of the potential in our study is expressed as $\Sigma_{Q \bar{Q}}\phi(r)$ (see Eq.(\ref{eq_Vcomplex})), where $\Sigma_{Q \bar{Q}}$ represents the 2-body selfenergy, and the interference function $\phi(r)$ is shown by the solid lines in Fig.~\ref{fig_phi}.

We compare our potential to that in Ref.~\cite{Bazavov:2023dci}, which has been extracted from lQCD data on WLCs using a Lorentzian spectral function with energy cutoff, as shown in Fig.~\ref{fig_Vcomp_WLC}, and to Ref.~\cite{Shi:2021qri} where it has been is fitted to lQCD results of bottomonium thermal widths using a Gaussian ansatz applied via deep neural networks, as shown in Fig.~\ref{fig_Vcomp_DNN}.
The real part of the potential in our study shows reasonable agreement with those presented in both Refs.~\cite{Bazavov:2023dci} and \cite{Shi:2021qri}. 
However, there is a notable difference in the imaginary part: in our study, it is approximately 2 to 3 times smaller than for the Lorentzian- and Gaussian-based fits at large distances, with even more significant differences at short distances. This observation aligns qualitatively with the findings in Appendix~\ref{sec_sfex2}: to describe the same lQCD data for bottomonium effective masses with extended operators, the Gaussian fit requires a much larger width than that obtained from the microscopic $T$-matrix calculations, as shown in Fig.~\ref{fig_SFbb_Tm_vs_GS_lattice}. Similarly, although the width required by the cut Lorentzian fit is smaller than that of the Gaussian fit, it remains larger than the width from the $T$-matrix. As indicated in Fig.~\ref{fig_Meff_Tm_vs_BW}, the  Lorentzian fit with cutoff still requires a larger width to achieve the necessary slope in $\Meff$ for reproducing the lattice data.
The comparison for the imaginary part of the potential can  probably be made more consistent if a more realistic spectral function were used in the lattice study.

\begin{figure*}[tbp]
\begin{minipage}[b]{1.0\linewidth}
\centering
\includegraphics[width=0.99\textwidth]{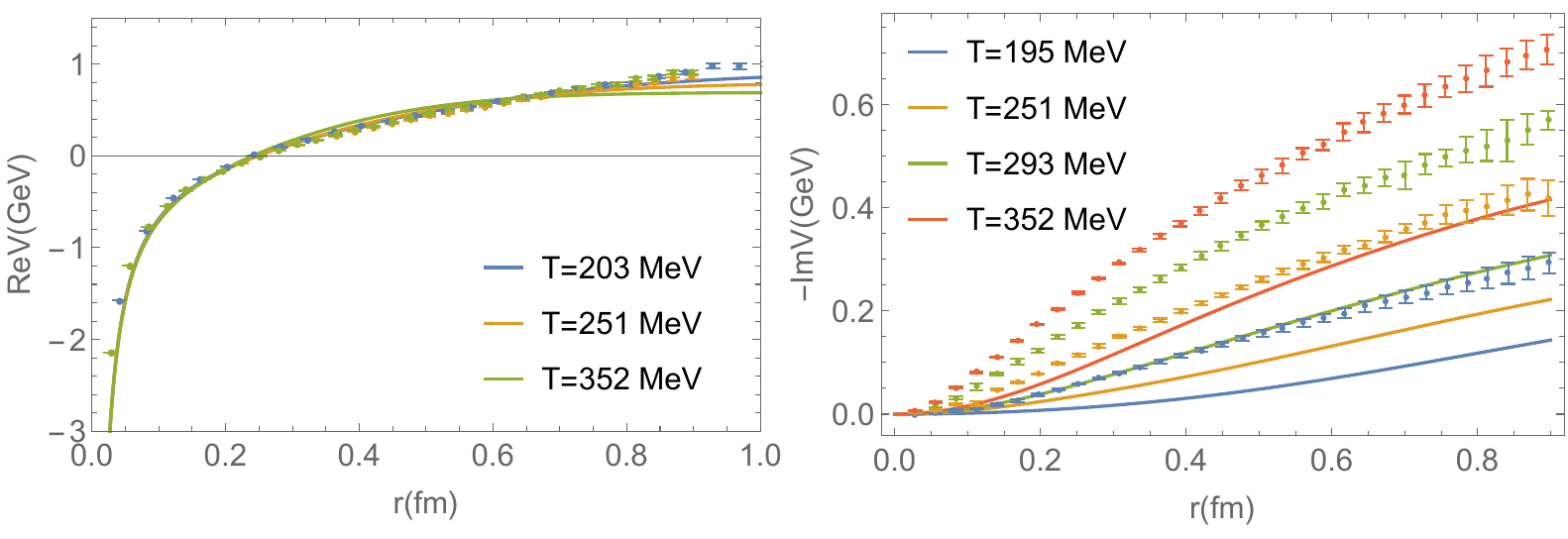}
\end{minipage}
\caption{Comparison of the real (left panel) and imaginary (right panel) parts of potential from the $T$-matrix approach (lines) with those from Ref.~\cite{Bazavov:2023dci} based on the Lorentzian fits with cutoff (dots with error bars).} 
\label{fig_Vcomp_WLC}
\end{figure*}

\begin{figure*}[tbp]
\begin{minipage}[b]{1.0\linewidth}
\centering
\includegraphics[width=0.99\textwidth]{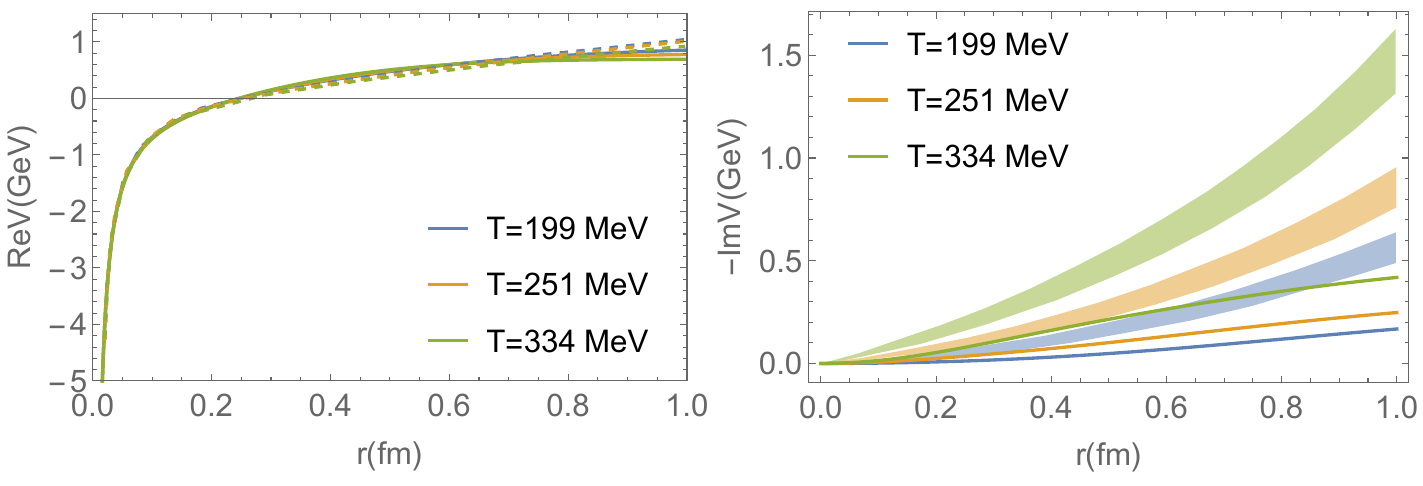}
\end{minipage}
\caption{Comparison of the real (left panel) and imaginary (right panel) parts of potential in the $T$-matrix approach (lines) with those from Ref.~\cite{Shi:2021qri} based on Gaussian fits (dashed lines for the real part and bands for the imaginary part).} 
\label{fig_Vcomp_DNN}
\end{figure*}

\bibliography{refnew}

\end{document}